\documentstyle{mn}
\def\bigstrut{\vrule width0pt height0.6truecm}
\font\japit = cmti10 at 11truept
\title[Linear power spectrum
of cosmological mass fluctuations]
{
\vglue-3.0truecm
\centerline{\japit Accepted for publication in Monthly Notices of the R.A.S.}
\vglue 2.5truecm
\noindent
Reconstructing the linear power spectrum
of cosmological mass fluctuations
}

\author[J.A. Peacock and S.J. Dodds]
{J.~A. Peacock$^1$ and S.~J. Dodds$^2$\\
$^1$Royal Observatory, \bigstrut Blackford Hill, Edinburgh EH9 3HJ, UK \\
$^2$Institute for Astronomy, University of Edinburgh,
Blackford Hill, Edinburgh EH9 3HJ, UK}

\def\japitem#1{\indent\llap{#1 $\;$}\hangindent\parindent}
\def\ref{\parskip =0pt\par\noindent\hangindent\parindent
    \parskip =2ex plus .5ex minus .1ex}
\def\gs{\mathrel{\raise1.16pt\hbox{$>$}\kern-7.0pt
\lower3.06pt\hbox{{$\scriptstyle \sim$}}}}
\def\ls{\mathrel{\raise1.16pt\hbox{$<$}\kern-7.0pt
\lower3.06pt\hbox{{$\scriptstyle \sim$}}}}
\def\pmb#1{\setbox0 =\hbox{#1}%
  \kern-.025em\copy0\kern-\wd0
  \kern.05em\copy0\kern-\wd0
  \kern-.025em\raise.0433em\box0 }

\def\erf{{\rm erf\,}}
\def\ss{\rm\scriptscriptstyle}

\def\A{&&\!\!\!\!\!\!\!\!\!\!\!}
\newcommand{\annrev} {ARA\&A}
\newcommand{\astast} {A\&A}

\newcommand{\apj} {ApJ}
\newcommand{\apjs} {ApJS}
\newcommand{\mn} {MNRAS}
\newcommand{\nat} {Nature}
\newcommand{\prl} {Phys. Rev. Lett.}
\newcommand{\be} {\begin{equation}}
\newcommand{\ee} {\end{equation}}
\newcommand{\ba} {\begin{eqnarray}}
\newcommand{\ea} {\end{eqnarray}}
\newcommand{\nn}{\nonumber \\}

\begin{document}

\maketitle

\begin{abstract}
We describe an attempt to reconstruct the initial conditions
for the formation of cosmological large-scale structure,
under the assumption of gravitational
instability in a Gaussian density field.
Information on the power spectrum of the primordial
fluctuations is provided by a variety of autocorrelation and
cross-correlation analyses on samples of different
classes of galaxy and galaxy clusters. These results
differ from the desired linear power spectrum because of
three modifying effects: bias, nonlinear evolution
and redshift-space distortions. We show how the latter
two effects can be corrected for analytically,
allowing the linear mass spectrum to be recovered
provided the bias is independent of scale for
a given class of galaxy. We argue that this is
a good assumption for large scales, which
is well verified in practice.

\hglue 1.0truecm
We apply this method to eight independent datasets,
and obtain excellent agreement in the estimated
linear power spectra for wavelengths
$\lambda\gs 10\,h^{-1}\rm Mpc$, given the following
conditions. First, the relative bias factors
for Abell clusters, radio galaxies, optical
galaxies and IRAS galaxies must be in the
ratios $b_{\ss A}:b_{\ss R}:b_{\ss O}:b_{\ss I} =
4.5:1.9:1.3:1$, to within 6 per cent rms.
Second, the data require a significant degree
of redshift-space distortion: $\Omega^{0.6}/b_{\ss I}
 = 1.0 \pm 0.2$. Third, low values of $\Omega$ and bias
are disfavoured because nonlinear evolution would
spoil the agreement in shape between galaxy and
cluster power spectra.
The amplitude of the preferred linear power
spectrum is only weakly dependent on $\Omega$
and agrees well at large wavelengths with
the normalization demanded by the COBE data
for a scale-invariant primordial spectrum,
provided $\Omega=1$ and gravity-wave anisotropies
are negligible. In this case, the shape of the spectrum
is extremely well described
by a CDM transfer function with an apparent value
of the fitting parameter $\Omega h =0.25$.
Tilted models, for which inflation requires a
large gravity-wave contribution to the COBE data,
predict too little power at 100 Mpc wavelengths.

\end{abstract}

\section{INTRODUCTION}

The simplest hypothesis for the origin of the
large-scale structure of the universe is that
it is the result of the operation of gravitational
instability on small initial density perturbations.
On grounds of economy, these are often
assumed to have the random-phase character common
in noise processes, and hence to form a Gaussian
random field. This has been the standard picture
for structure formation for the half century since
the pioneering studies of Lifshitz, given added
motivation more recently by inflationary theories in which the
initial perturbations are supplied by quantum
fluctuations at early times.

If this picture is correct, the only quantity
needed for a complete statistical description
of the cosmological density field is the
power spectrum of the fluctuations at some early
time (or its linear-theory extrapolation to the present).
Observationally, much progress has been
made in recent years towards the goal of determining
the power spectrum, fulfilling the programme
outlined by Peebles (1973).
New generations of deep redshift surveys
have allowed the clustering
of various classes of galaxy to be determined up to
the contribution from wavelengths of several hundred
Mpc. In parallel, new analysis techniques have been
developed in order to extract the long-wavelength portion
of the power spectrum more sensitively (e.g.
Feldman, Kaiser \& Peacock 1993 [FKP] and refs therein).

The intention of this paper is to compare
various recent determinations of galaxy clustering, and to
see if there exists a single consistent picture
for the underlying mass fluctuations.
It is an updated version of a previous attempt
in this direction (Peacock 1991), but with
several important improvements in addition to a
great increase in data.
In essence, there are three filters that cause the
observed clustering properties of galaxies to depart from the
desired linear mass power spectrum:

\japitem{(i)}Nonlinear evolution. On small scales, perturbation
theory fails and the mass power spectrum departs in a
complicated way from a linear extrapolation of the initial
conditions.

\japitem{(ii)}Redshift-space effects. Because 3D datasets use
redshift as a radial coordinate, the apparent density field that
results is distorted through the existence of peculiar velocities.
Even for perfect data, the redshift-space power spectrum is
not the same as that in real space.

\japitem{(iii)}Bias. The fact that different species of galaxy
follow the mass distribution with different degrees of fidelity
is a major problem in relating observations to theory. To correct
for bias in principle requires a detailed model for
how the effect arises.

None of these effects have been handled very thoroughly in previous
work. The issue of bias is the most difficult, and is really
only tractable on large scales where the degree of
bias can be assumed constant. It is now possible to have
a better idea of where this approximation is valid,
and we discuss this issue in Section 2.
Previously, nonlinear distortions were either ignored
or treated by comparing
nonlinear data with an evolved $N$-body model.
However, thanks to the insight of Hamilton et al.
(1991; HKLM), it is possible to correct the data for the
effects of nonlinearities. We discuss their method
in Section 3 and give a number of generalizations.
Redshift-space distortions have usually been treated
by a simple scaling of amplitude analyzed by Kaiser (1987), but
this is inapplicable on small scales. We give an improved analysis
in Section 4.

Given a method for treating the practical distortions
of power spectra, there are two possible approaches.
There is an honourable tradition which states that
it is better to apply any corrections to the theory under test,
and to compare the modified model with the raw data.
Nevertheless, we shall do the opposite and
estimate the linear spectrum by correcting
the data. This has two advantages: no model is
needed, and the power spectrum can be found empirically;
by comparing the various estimates, we can then see directly
if all datasets are consistent with each other.
In Section 5, we assemble the most recent power-spectrum
data and apply the above tools to deduce the linear
power spectrum. This empirical reconstruction is
compared with a variety of {\it a priori\/} models
in Section 6, and the main points of the paper
are summarized in Section 7.

\section{GALAXY AND CLUSTER CORRELATIONS IN GAUSSIAN MODELS}

\subsection{Evidence for Gaussian fluctuations}

Since a good part of the analysis in this paper rests on
the assumption of a Gaussian density field, we should
start by considering the evidence that this is a good
approximation.

The evidence has to be gathered on large scales, because
nonlinear evolution inevitably induced non-Gaussian
statistics on small scales, whatever the initial
statistics. The most direct test was carried out by
FKP, who looked at the
distribution of power measured for individual modes
in a power-spectrum analysis of the IRAS QDOT redshift
survey. For a Gaussian field, such modes should
have power values that are independently exponentially
distributed. This was found to be the case out
the limit of the statistics -- powers of about 10 times the mean.
This is not a complete test of the Gaussian hypothesis:
it is equivalent to asking in real space whether the
one-point density distribution is Gaussian. Further
information is provided by higher-order $k$-space
correlations which test for independence of the modes.
Nevertheless, it is worth recalling that there have been
suggestions that even this lowest-order test is badly
violated. On the basis of a pencil-beam redshift
survey, Broadhurst et al. (1990) and Szalay et al. (1991)
have suggested that there is gross non-Gaussian
behaviour on large scales, based on the
existence of strong quasi-periodic power at a few wavelengths.
There is no need to repeat here the counter-arguments
given by Kaiser \& Peacock (1990); it should suffice to note
that the QDOT sample is deep enough that it encompasses
several of the suggested periods in a large number of
independent directions, yet no non-Gaussian signature
is detected.

Any initial Gaussian nature of the field is completely
erased on very small scales, but on intermediate scales
the field develops a skewness which can be analyzed
perturbatively (Peebles 1980). The observed degree
of skewness appears to be in accord with this prediction
(Gazta\~naga 1992; Bouchet et al. 1993), which gives
further support to the Gaussian hypothesis.
This is not a definitive test, since most bias mechanisms
will induce skewness; what is observed is a mixture of
this effect with primordial skewness, plus the effects
of gravitational evolution. Nevertheless, simple Gaussian models
without a strong degree of bias do account for the data well.

A variety of other tests have been suggested, including
the topology of isodensity surfaces (Hamilton, Gott
\& Weinberg 1986; Coles \& Plionis 1991; Moore et al. 1992)
and the one-point distribution of the velocity field
(Nusser \& Dekel 1993; Kofman et al. 1993).
It is fair to say that none of these methods has
produced any evidence against primordial Gaussian
statistics. However, as usual in statistics, it is
necessary to choose the null hypothesis with care.
It is certainly the case that not all tests are
necessarily very powerful; the central limit
theorem means that a variety of non-Gaussian
processes may yield nearly Gaussian behaviour in
experiments where limited resolution averages over
different regions of space (Scherrer 1992).
Thus, some of the more popular models based on
topological singularities (strings, textures etc.)
may still be allowed by existing data (e.g.
Gooding et al. 1992). It
will however be interesting to see such theories
confronted with the FKP result,
particularly as such statistics will become more
demanding as datasets increase.

For the present, it is enough to note that there
is empirical reason to believe that the statistics
of the large-scale density field are close to Gaussian.
If this is so, then there are consequences for
the clustering of galaxy systems, as discussed
below. As we will see, these predictions are
verified in practice, which is one further
piece of supporting evidence for the Gaussian picture.

\subsection{Bias in galaxy and cluster correlations}

In a Gaussian model, the correlations of different
classes of galaxy system can be directly related
to the underlying density field, with the power
spectra being proportional on large scales
\be
\Delta^2(k) =b^2\,\Delta^2_{\rm mass}(k).
\ee
Here and below, we shall use a dimensionless
notation for the power spectrum designed to
minimise uncertainties from differing Fourier
conventions. In words, $\Delta^2$ is the
contribution to the fractional density variance
per bin of $\ln k$; in the convention of
Peebles (1980), this is
\be
\Delta^2(k)\equiv{d\sigma^2\over d\ln k} ={V\over (2\pi)^3}
\, 4\pi \,k^3\, |\delta_k|^2.
\ee
The justification for the above relation is the assumption,
introduced by Kaiser (1984), Peacock \& Heavens (1985)
and Bardeen et al. (1986; BBKS),
that the sites of massive objects such as clusters
can be identified at early times as high
peaks in the linear density field. Such a
scheme might be termed Lagrangian bias, and $b$ is
called a bias parameter. This is rather sloppy:
biased galaxy formation usually means
the situation where light does not trace
mass in the universe, but clusters would
still be more correlated than the mass even
if the galaxy distribution followed the mass exactly.
However, this useage is too firmly embedded in
the literature to make it worth fighting; we
will therefore describe the enhanced correlations
of clusters as bias.

As the density field evolves, the initial
statistical clustering in Lagrangian space is supplemented as
dynamics moves objects from their initial sites.
Owing to the equivalence principle, all
objects move in the same way, so that the
overall observed clustering in Eulerian space is
\be
1+\delta_{\rm Euler} =(1+\delta_{\rm Lagrange})\,(1+\delta_{\rm dynamics}).
\ee
In the linear regime, we therefore have
\be
\delta_{\rm Lagrange} =(b-1)\delta_{\rm dynamics}.
\ee
Bond \& Couchman (1988) showed how this decomposition
could be used to compute exact total
correlations, under the assumption that the
dynamical evolution obeyed the Zeldovich (1970) approximation.
Mann, Heavens \& Peacock (1993) applied this method
to the calculation of cluster correlations. In
practice, the statistical contribution tends
to dominate for scales larger than the filter
size used to define clusters (a few Mpc); the
cluster distribution has not undergone strong
dynamical evolution, and most clusters are
close to their original sites.

Although the idea of Lagrangian bias was borrowed
by BBKS from its cluster origins and applied as
a model for biased galaxy formation, it
may be more fruitful to think of galaxy
bias in a purely Eulerian way, where
the density of galaxies is some function of
the final mass density. This has long been
advocated by Einasto and collaborators, with
galaxy formation being suppressed in low-density
regions (Einasto, Joeveer \& Saar 1980).
More recently, studies of the
operation of dissipation in numerical simulations
has produced a more direct physical justification
for relating the galaxy and mass density
fields through a single nonlinear function
(Cen \& Ostriker 1992).

These contrasting views of the origin of
cluster and galaxy bias lead to rather different
approaches when attempting to use
clustering data to infer the mass fluctuations.
For clusters, the statistical bias is dominant,
and we may assume that the clusters reflect
mainly the initial conditions. Conversely,
it is reasonable to believe that galaxies
come close to tracing the mass. Many studies
have indicated that different classes of
galaxy follow the same `skeleton' of voids,
filaments, walls \& clusters, while differing
most markedly in regions of high density
(e.g. Babul \& Postman 1990, Strauss et al. 1992). This last
effect may not be so important: despite having
densities differing by factors of close to 10
in rich clusters, we shall see below that
IRAS and optical galaxies have bias factors
within about 30 per cent of each other.
This is analogous to the findings of Cen \&
Ostriker (1992): even though their model
has a highly nonlinear dependence of galaxy
density on mass density for high densities,
the power spectra are proportional on
most scales, even down to the point where
$\Delta^2(k)\sim1$.
In any case, it is important to keep in mind
that we are not interested in exactly how
a given class of galaxy does or does not follow the
density field: a variety of different
bias schemes could give the same galaxy
power spectrum, even though the light
distributions would be model dependent.

The above discussion motivates the assumptions
that we shall use below to make estimates of
the linear power spectrum. We shall adopt the
extreme approximations that the cluster
distribution contains information only about the
linear power spectrum, whereas the galaxy
distribution mainly measures the nonlinear
density field:
\ba
\Delta^2_{\ss C} = \A b_{\ss C}^2\Delta^2_{\ss L} \\
\Delta^2_{\ss G} = \A b_{\ss G}^2\Delta^2_{\ss NL}
\ea
A further way of understanding this distinction is
to consider the following illustrative model:
populate the universe with identical spherical
protocluster perturbations. At some critical time,
these will turn round and virialize, producing
a large excess of small-scale power in the nonlinear
density field. However, at this time, the cluster
centres will still be weakly perturbed: the existence
of the small-scale power is what allows us to say that
clusters are present, but there is no reason to
expect this power to manifest itself in many close pairs
of cluster centres. Ultimately, our hypothesis must
submit to the test of numerical simulation, but
for the present it should certainly be closer to the truth to
say that clusters respond to the linear power
spectrum, rather than to the nonlinear one.

Although the above bias factors are calculable
given a specific bias model, we shall treat
them as unknowns to be determined from the data.
It is clear that the assumption of constant
bias factors cannot be exact,
and will certainly break down at small scales.
To some extent, the domain of validity
can be found empirically, by seeing whether
it is possible to make a consistent picture
in this way from all the available data.
In practice, we shall use
data at wavenumbers $k\ls 0.6\,h\,\rm Mpc^{-1}$,
i.e. wavelengths $\lambda\gs 10\,h^{-1}\rm Mpc$
(as usual, $h\equiv H_0/100\,\rm km\,s^{-1}Mpc^{-1}$),
so we are only dealing with the large-scale
mass distribution.

There is a third way in which the mass power
spectrum may be inferred, which is to use
cross-correlation data from two catalogues,
in addition to the respective autocorrelations.
In this case, it is not so obvious whether
we measure more nearly the linear or nonlinear
correlations. In practice, we shall use data
on large enough scales that the distinction
will not be so important; we therefore assume
a relation to linear theory
\be
\Delta^2_{\ss CG} = b_{\ss C}b_{\ss G}\Delta^2_{\ss L}.
\ee
This provides a useful consistency test of
our assumptions: the cluster-galaxy cross-correlation
should be the geometrical mean
of the separate auto-correlations.

\section{NONLINEAR EVOLUTION OF POWER SPECTRA}

To implement the above assumptions requires some
way of relating linear and nonlinear power spectra;
until recently, this would have required
$N$-body modelling. However, in a marvelous
piece of alchemy, Hamilton et al. (1991; HKLM)
gave a universal analytical formula for accomplishing
the linear $\leftrightarrow$ nonlinear mapping.
The conceptual basis of their method can be
understood with reference to the spherical
collapse model. For $\Omega =1$ (the only case
they considered), a spherical clump
virializes at a density contrast of order
100 when the linear contrast is of order unity.
The trick now is to think about the density
contrast in two distinct ways. To make a
connection with the statistics of the density field,
the correlation function $\xi(r)$ may be
taken as giving a typical clump profile.
What matters for collapse is that the integrated
overdensity reaches a critical value, so one should
work with the volume-averaged correlation function
$\bar\xi(r)$. A density contrast of
$1+\delta$ can also be thought of as arising
through collapse by a factor $(1+\delta)^{1/3}$ in radius,
which suggests that a given non-linear correlation
$\bar\xi_{\ss NL}(r_{\ss NL})$ should be thought
of as resulting from linear correlations on a
linear scale
\be
r_{\ss L} =[1+\bar\xi_{\ss NL}(r_{\ss NL})]^{1/3} r_{\ss NL}.
\ee
This is one part of the HKLM procedure.
The second part, having translated scales as above,
is to conjecture that the nonlinear correlations are
a universal function of the linear ones:
\be
\bar\xi_{\ss NL} (r_{\ss NL}) = f_{\ss NL}[\bar\xi_{\ss L}(r_{\ss L})].
\ee
The asymptotics of the function can be deduced
readily. For small arguments $x\ll1$, $f_{\ss NL}(x) \simeq x$;
the spherical collapse argument suggests
$f_{\ss NL}(1)\simeq 10^2$.
Following collapse, $\bar\xi_{\ss NL}$ depends on scale
factor as $a^3$ (stable clustering), whereas
$\bar\xi_{\ss L}\propto a^2$; the large-$x$ limit is therefore
$f_{\ss NL}(x)\propto x^{3/2}$. HKLM deduced from numerical
experiments that the exact coefficient was
\be
f_{\ss NL}(x)\rightarrow 11.68 \,x^{3/2}
\ee
and obtained a numerical fit that interpolated
between these two regimes, in a manner that
empirically showed negligible dependence on
power spectrum.

To use this method in the present application,
we need two generalizations: we need to make the
method work with power spectra, and we need
the analogous results with $\Omega\ne 1$.
In principle, the translation between $\bar\xi(r)$
and $\Delta^2(k)$ is straightforward, but
obtaining stable numerical results is not so
easy. One route is to use the relations
between $\bar\xi(r)$ and $\xi(r)$
\ba
\bar\xi(r) = \A {3\over r^3}\int_0^r \xi(x)\; x^2\;dx \\
\xi(r)    = \A {d\, [r^3\, \bar\xi(r)]\over d[r^3]},
\ea
followed by the Fourier relations between $\xi(r)$
and $\Delta^2(k)$
\ba
\xi(r) = \A \int_0^\infty\Delta^2(k)\,{\sin kr\over kr}\; {dk\over k} \\
\Delta^2(k) = \A {2k^3\over\pi}
    \int_0^\infty\xi(r)\,{\sin kr\over kr}\; r^2\;dr.
\ea
This approach is not so attractive. To obtain the
nonlinear power spectrum from the linear one
requires two numerical integrations, differentiation,
followed by one further integration. It is
possible to do a little better by manipulating
the above equations to relate $\Delta^2(k)$ and $\bar\xi(r)$
directly
\ba
\bar\xi(r) = \A\int_0^\infty\Delta^2(k)\;{dk\over k}\;{3\over (kr)^3}
[\sin kr- kr \cos kr] \\
\Delta^2(k) = \A{2k^3\over 3\pi}\int_0^\infty\bar\xi(r)\, r^2\,dr
\;{1\over (kr)} [\sin kr- kr \cos kr],
\ea
where the last relation holds provided that $\bar\xi(r)\rightarrow 0$
faster than $r^{-2}$ at large $r$ (i.e. a spectrum
which asymptotically has $n>-1$, a valid assumption
for spectra of practical interest).
This looks better, since there are now only two integrations
required, and furthermore efficient methods
exist for dealing with integrations with
sin and cos weightings in the integrand.
However, because the window function consists
of the difference of two such terms, life is still
not so easy: evaluating the two parts of the
integral separately gives a result as a
difference of two large numbers, which is
thus generally of low accuracy.
The most satisfactory practical procedure seems
to be a mixture of the two possibilities:
(i) evaluate a table of $\bar\xi_{\ss L}(r)$ values for a given
linear power spectrum by evaluating the
oscillatory integral directly; (ii) transform to
a table of $\bar\xi_{\ss NL}(r)$ values using the
HKLM procedure; (iii) fit splines to the result
and differentiate to get $\xi_{\ss NL}(r)$;
(iv) Fourier transform to get $\Delta^2_{\ss NL}(k)$.
The accuracy of the result can be improved in
the final step by transforming $\xi_{\ss NL}(r)-\xi_{\ss L}(r)$,
which vanishes rapidly at large $r$, and then adding
$\Delta^2_{\ss L}(k)$ to the answer.

The above process is still rather time-consuming and
inelegant; it would be much better to make
the HKLM method work directly in terms of
power spectra, and this is usually possible.
The main idea is that $\bar\xi(r)$ can often
be thought of as measuring the power at some
effective wavenumber: it is obtained as
an integral of the product of $\Delta^2(k)$,
which is often a rapidly rising function, and
a window function which cuts off rapidly at high $k$.
The answer can be approximated by replacing the
exact window function by the Gaussian which
is equivalent to second order in  $k$:
\ba
\bar\xi(r) = \A\Delta^2(k_{\rm eff}) \\
k_{\rm eff} = \A\left[{([n+1]/2)\,!\over 2}\right]^{1/(n+3)}
{\sqrt{10}\over r},
\ea
where $n$ is the effective power-law index of the power
spectrum. This approximation is within a few per cent of
the exact integration provided $n\ls 0$.
The effective wavenumber is insensitive to $n$,
and is within 20 per cent of $2.4/r$ over the
range $-2<n<0$.
In most circumstances, it is therefore an excellent
approximation to use the HKLM formulae directly to
scale wavenumbers and powers:
\ba
\A \Delta^2_{\ss NL}(k_{\ss NL}) = f_{\ss NL}
[\Delta^2_{\ss L}(k_{\ss L})] \\
\A k_{\ss L} = [1+\Delta^2_{\ss NL}(k_{\ss NL})]^{-1/3} k_{\ss NL}.
\ea
Even better, it is not
necessary that the number relating $1/r$ and
$k_{\rm eff}$ be a constant over the whole
spectrum. All that matters is that the number
can be treated as constant over the limited range
$r_{\ss NL}$ to $r_{\ss L}$. This means that
the deviations of the above formulae from the
exact transformation of the HKLM procedure
are only noticeable in cases where the
power spectrum deviates markedly from a
smooth monotonic function, or where either
the linear or nonlinear
spectra are very flat ($n\ls -2$).
Even this is not obviously a problem,
since the HKLM procedure itself is
not exact and does not work so well for
flat spectra, $n\ls -2$ (A.S. Hamilton, private communication).
Our approximation is illustrated
in Figure 1, which shows the result of
nonlinear evolution on spectra with and without
a short-wavelength cutoff. Whether the HKLM
method actually applies to the first situation
is an interesting question which we hope
to investigate elsewhere. Other cases where
our approximation would fail include
power spectra with the oscillations characteristic
of pure baryon models. However, since the
data studied here reveal no trace of such
sharp features in the power spectrum, we
may use the direct approximation for the
nonlinear evolution of the power spectrum
with confidence.

It remains to generalise the result from the
$\Omega =1$ model considered by HKLM. This can
be done partly analytically. The argument that leads
to the $f_{\ss NL}(x)\propto x^{3/2}$ asymptote in the
nonlinear transformation is just that linear
and nonlinear correlations behave as $a^2$ and $a^3$
respectively following collapse. If collapse
occurs at high redshift, then $\Omega =1$ may be
assumed at that time, and the nonlinear correlations still
obey the $a^3$ scaling to low redshift. All that
has changed is that the linear growth is
suppressed by some $\Omega$-dependent factor $g(\Omega)$.
It then follows that the large-$x$ asymptote of
the nonlinear function is
\be
f_{\ss NL}(x)\rightarrow 11.68 \,[g(\Omega)]^{-3}\,x^{3/2}.
\ee
According to Carroll, Press \& Turner (1992), the
growth factor
may be approximated almost exactly by
\be
g(\Omega) =\frac{5}{2}\Omega_m\left[\Omega_m^{4/7}-\Omega_v+
(1+\Omega_m/2)(1+\Omega_v/70)\right]^{-1},
\ee
where we have distinguished matter ($m$) and vacuum
($v$) contributions to the density parameter explicitly.
We shall generally use $\Omega$ without a subscript
to mean $\Omega_m$ hereafter.

To interpolate between the expected nonlinear asymptote and
the linear regime, numerical experiments are necessary.
We therefore wrote a PM $N$-body code, which was
used to evolve a variety of initial
spectra to a final state of given $\Omega_m$ and $\Omega_v$;
typically $64^3$ particles and a $128^3$ mesh were
used. At a later stage of the investigation, we
were able to check our results with the superior
resolution provided by the AP$^3$M code of Couchman (1991).
It was also possible to compare with
low-density CDM models published by Davis et al. (1985)
and Kauffmann \& White (1992).
Our conclusion is that a near-universal behaviour analogous
to that of HKLM does appear to exist for low-density models,
at least for the linear spectra with power-law indices
$-2<n<0$ that we were able to test.
We have produced the following fitting formula for the
generalized $f_{\ss NL}$. This is designed to match
the HKLM expression almost exactly in the $\Omega=1$
limit, and to describe the main features of the
alterations encountered in low-density models.
The accuracy is approximately 10 per cent in terms of
the deduced linear power
$\Delta^2_{\ss L}$ corresponding to a given $\Delta^2_{\ss NL}$,
over the range $0.3\ls g(\Omega) \ls 1$.
\be
f_{\ss NL}(x) =x \; \left[{ 1+0.2\beta x +(A x)^{\alpha\beta} \over
1 + ([A x]^\alpha g^3(\Omega)/[11.68 x^{1/2}])^\beta}\right]^{1/\beta},
\ee
where the parameters are $A = 0.84 [g(\Omega)]^{0.2}$,
$\alpha= 2/[g(\Omega)]$,
and $\beta=2g(\Omega)$.
This fit says that the transition region between the linear
and nonlinear regimes is dominated by an $f_{\ss NL}\propto x^{1+\alpha}$
power law, which becomes very steep for low-density
models. This steepening has long been familiar from
$N$-body models, and the apparent power-law nature
of the spectrum can been used as an argument
against low-density models. We shall end up
making a rather similar argument here.

It is useful to have an analytical expression
for the inverse function, and the following
agrees with the exact inverse of our formula
to within a typical maximum
error of a few per cent over the range of interest:
\be
f^{-1}_{\ss NL}(y) =y \; \left[{
1 + (B y^{\gamma-1/3} [g^3(\Omega)/11.68]^{2/3})^\delta
\over
1+ 0.2\delta y + (B y^\gamma)^\delta
}\right]^{1/\delta},
\ee
where $B=0.96 [g(\Omega)]^{0.07}$, $\gamma=1.03 - 0.39[g(\Omega)]^{0.5}$,
and $\delta=5 [g(\Omega)]^{0.3}$.
A plot of the $\Omega$-dependent nonlinear function is
shown in Figure 2. We show only models with zero vacuum energy,
since the above reasoning shows that all that matters
is the linear growth-suppression factor $g(\Omega)$.
Note that, for spatially flat vacuum-dominated models,
the growth suppression is rather more modest (roughly
$g(\Omega) =\Omega^{0.2}$) than in models with
zero vacuum energy (roughly $g(\Omega) =\Omega^{0.7}$).
Our results (and those of HKLM) apply only to initial
conditions with Gaussian statistics. It is an interesting
question to what extent the method will also apply
to non-Gaussian models, an we hope to investigate
this elsewhere. Some idea of the likely degree of universality
may be gained from the non-Gaussian models studied
by Weinberg \& Cole (1992). They found that the
nonlinear power spectrum was very similar for a
range of initial models, with the exception only
of those which were strongly skew-negative. It therefore
seems likely that mildly non-Gaussian models
such as cosmic strings should be treated
correctly by the method we have given.

We now have the required means of deducing the initial
conditions that correspond to a given observed
non-linear mass spectrum. As an example, we show
in Figure 3 the initial conditions required to
create the canonical correlation function $\xi(r)=(r/r_0)^{1.8}$
-- i.e. $\Delta^2(k)=(k/k_c)^{1.8}$, where $k_c=1.058/r_0$.
For low-density models, the initial conditions
require an enormous `bite' to be taken out of the
spectrum for $k$ between $k_c$ and several times
$k_c$. The reconstructed spectrum tends to be
very flat and close to $\Delta^2(k)=1$ over a large
range of wavenumber. Conversely, for $\Omega=1$
the effects of nonlinearities are not very severe in
this case until we reach $\Delta^2(k)\gs 10$.
These differences will be important when we come to
linearize the observed data.

\section{REDSHIFT-SPACE DISTORTIONS}

With the exception of surveys where angular
data are deprojected to obtain an estimate of
the spatial power spectrum, three-dimensional
clustering data generally involve redshift surveys
where the radii are distorted by peculiar
velocities.
There are two effects to consider. On large
scales, a linear analysis should be valid and
we have the anisotropic effect noted by Kaiser (1987):
\be
\delta_k\rightarrow b\,\delta_k\, (1+f\mu^2/b),
\ee
where $\mu$ is the cosine of the angle between the wavevector
and the line of sight.
The function $f(\Omega)\simeq \Omega^{0.6}$ is the
well-known velocity-suppression factor due
to Peebles, which is in practice a function
of $\Omega_m$ only, with negligible dependence
on the vacuum density (Lahav et al. 1991).
The anisotropy arises because mass flows
from low-density regions onto high density sheets,
and the apparent density contrast of the
pattern is thus enhanced in
redshift space if the sheets lie near the plane
of the sky. If we average this anisotropic effect
by integrating over a uniform distribution of $\mu$,
the net boost to the power spectrum is
\be
|\delta_k|^2 \rightarrow b^2\,|\delta_k|^2\;
\left(1+{2\over3}[f/b] + {1\over 5}[f/b]^2\right).
\ee

On small scales, this is not valid. The main effect
here is to reduce power through the radial smearing
due to virialized motions  and the associated
`finger-of-God' effect. This is hard to treat
exactly because of the small-scale velocity correlations.
A simplified model was introduced by Peacock (1992)
in which the small-scale velocity field is taken
to be an incoherent Gaussian scatter with 1D rms
dispersion $\sigma$. This turns out to be quite a
reasonable approximation, because the observed pairwise
velocity dispersion is a very slow function of separation,
and is all the better if
the redshift data are afflicted by significant
measurement errors (which should be included in $\sigma$).
This model is just a radial convolution,
and so the $k$-space effect is
\be
\delta_k\rightarrow \delta_k\, \exp[-k^2\mu^2\sigma^2/2].
\ee
This effect in isolation gives an average isotropic factor of
\be
|\delta_k|^2 \rightarrow |\delta_k|^2\; {\sqrt{\pi}\over2}
\, {\erf(k\sigma)\over k\sigma}
\ee
and produces only mild damping (one power of $k$ at
large $k$).

Some workers (e.g. Fisher et al. 1992; Kofman, Gnedin \& Bahcall
1993) have combined
the above two effects simply by multiplying the
two power correction factors to achieve a total
distortion. However, this is not correct: both
terms are anisotropic in $k$ space and they interfere
before averaging: $\langle A^2B^2\rangle\ne
\langle A^2 \rangle\langle B^2 \rangle$.
For the present paper, it is also interesting
to consider the case of cross-correlation
where each of two catalogues gives a different
measure of the same underlying density field.
The model for the effect in $k$ space of cross-correlation is
then the product of two separate factors of the
above form
\ba
|\delta_k|^2 \rightarrow \A  b_1b_2\, |\delta_k|^2
(1+f\mu^2/b_1)(1+f\mu^2/b_2) \times \nn
\A \exp[-k^2\mu^2(\sigma_1^2+\sigma_2^2)/2].
\ea
The overall effect is obtained by averaging over $\mu$, and
looks more complicated than it really is:
\be
|\delta_k|^2 \rightarrow b_1b_2\,|\delta_k|^2\; G(y,\alpha_1,\alpha_2),
\ee
where
\ba
\A y^2 \equiv k^2(\sigma_1^2+\sigma_2^2)/2 \\
\A \alpha \equiv f(\Omega)/b \\
\A G(y,\alpha_1,\alpha_2) = \nn
\A \quad {\sqrt{\pi}\over 8}\,  {\erf\,y\over y^5}
\,[3\alpha_1\alpha_2+2(\alpha_1+\alpha_2)y^2+4y^4] \nn
\A \quad -\; {\exp -y^2\over 4y^4}\,
[\alpha_1\alpha_2(3+2y^2)+2(\alpha_1+\alpha_2)y^2].
\ea
This simplifies a little in the case of autocorrelations,
where indices 1 and 2 are equivalent.
The interesting aspect of this formula is that
the linear boost is lost at large
$k$, where the result is independent of $\Omega$
(as is obvious from the anisotropic form: the main contribution
at large $k$ comes from small $\mu$).
The true damping at large $k$ is thus more severe than
would be obtained by multiplying the power corrections
prior to angular averaging. The simulations of
Gramann, Cen \& Bahcall (1993) show a good level of
agreement with the above formula in the autocorrelation case.
The result is reassuringly insensitive to the assumed form
for the small-scale velocity distribution function;
if we take an exponential instead of a Gaussian, we
find the same result at small $k$:
\ba
G(y,\alpha_1,\alpha_2) \simeq
\A \left(1+{\alpha_1+\alpha_2\over 3}+{\alpha_1\alpha_2\over5}\right) \nn
\A- \left({1\over 3} +
{\alpha_1+\alpha_2\over 5}+{\alpha_1\alpha_2\over7}\right) \; y^2,
\ea
and the large-$y$ limit becomes $G\rightarrow \pi/(2^{3/2}y)$ instead
of $G\rightarrow \pi^{1/2}/(2y)$.

In practice, the relevant value of $\sigma$ to choose is
approximately $1/\sqrt{2}$ times the pairwise dispersion $\sigma_\parallel$
seen in galaxy redshift surveys. According to the most
recent compilation of velocity results by Mo, Jing \& B\"orner (1993),
this corresponds to the figure (adopted hereafter) of
\be
\sigma\simeq 300\;{\rm kms^{-1}}.
\ee
To this, we should add in quadrature any errors in measured
velocities. The relatively low value of this dispersion is
of course a significant problem for some high-density
models. Gramann, Cen \& Bahcall (1993) argue that redshift-space
power spectra of CDM models fit observation very well, mainly
because the predicted pairwise dispersion is so high in these models.
As we shall see below, such an unrealistically  large dispersion would spoil
the agreement between datasets in real and in redshift space.

\section{POWER-SPECTRUM RECONSTRUCTION}

\subsection{Data}

We now apply the above tools to some of the more recent
results on the clustering power spectrum. We shall
consider eight distinct sets of data, which fall
into several distinct classes:

\japitem{(i)}Real-space clustering of galaxies. Baugh \& Efstathiou
(1993) have applied a deprojection procedure to the angular
clustering of the APM galaxy survey to infer the
nonlinear power spectrum of optically-selected galaxies
without redshift-space distortions.
This paper considers the large-scale power spectrum, and we
have thus used the APM data at $k<1\,h\,\rm Mpc^{-1}$ only.
To allow comparison with other datasets, we have also
set a lower limit of $k>0.015\,h\,\rm Mpc^{-1}$.

\japitem{(ii)}Redshift-space clustering of galaxies.
We consider three datasets: FKP
for IRAS galaxies (the QDOT sample); Loveday et al.
(1993) for the Stromlo/APM survey; Vogeley et al.
(1992) for the CfA survey. The last paper quotes results
for two separate subsets; we have adopted a straight
mean of the two sets of data.
We have not used the IRAS data of Fisher et al. (1992),
which is systematically lower than that of FKP.
As discussed by FKP, this seems most likely to be a
local sampling effect. In any case, it is the deeper
QDOT sample used by FKP which also appears in
cross-correlation analyses (see [iv] below).

\japitem{(iii)}Redshift-space clustering of groups
and clusters of galaxies. We use the power spectrum
for $R\ge 1$ Abell clusters from Peacock \& West (1992)
and also radio galaxies from Peacock \& Nicholson (1991),
on the assumption that the strongly enhanced clustering
of these latter objects may be attributed to their location
in moderately rich environments.

\japitem{(iv)}We also use the cross-correlation between
IRAS galaxies and Abell clusters or radio galaxies
from Mo, Peacock \& Xia (1993).

Not all of the above data are available directly in power spectrum
form. In cases where what is published is a cell variance
or a measure of $\bar\xi(r)$, we have used the notion of
an effective wavenumber, as discussed above and in Peacock (1991).
The treatment of errors requires some discussion.
Only FKP give a full realistic
error covariance matrix for their data; the other
datasets give errors ranging from Poisson estimates
to field-to-field errors, but with no discussion of
the independence of the measurements at different $k$.
For consistency, we have therefore used a fraction of
the FKP data, spaced widely enough
to be roughly independent. Any imprecision in this procedure,
plus unrecognised systematics, will become apparent
once the various datasets are compared with each other.

The raw power-spectrum data are plotted in Figure 4.
There is a wide range of power measured, ranging
over perhaps a factor 20 between the real-space
APM galaxies and the rich Abell clusters. We
now have to see to what extent these measurements
are all consistent with one Gaussian power spectrum
for mass fluctuations.

\subsection{Implications for bias and $\Omega$}

The reconstruction analysis has available eight
datasets containing 91 distinct $k-\Delta^2$ pairs.
The modelling has available five free parameters
in the form of $\Omega$ and the four bias parameters
for Abell clusters, radio galaxies, optical
galaxies and IRAS galaxies ($b_{\ss A}$, $b_{\ss R}$,
$b_{\ss O}$, $b_{\ss I}$). We optimized the model
by making independent determinations of $\Delta^2_{\ss L}(k)$
for each dataset and then comparing them. This
was done in practice by dividing the range $0.01<k<0.1$
$h\,\rm Mpc^{-1}$ into 20 bins, evaluating a weighted
mean power and a $\chi^2$ for each bin. The likelihood
of the model is given in terms of the summed $\chi^2$
values:
\be
{\cal L}\propto \exp -\chi^2/2.
\ee
At this stage, the question arises of whether the
errors are realistic, which may be judged from whether
the overall $\chi^2$ matches the number of degrees of freedom:
in fact, it does not. A procedure which ensures the
required match is to add some constant rms error $\epsilon$ in
quadrature to the existing errors. In practice
\be
\epsilon=23\ \rm per\ cent
\ee
is required for the best fitting model. Such a fudge is
unsatisfactory and indicates a failure of understanding
of the data errors. However, there are grounds
for suspecting that some of the published errors are
too low, so $\epsilon$ is not a surprisingly large
correction. There may be excessive democracy here, in
that the formally most accurate datasets are penalized
most strongly by this procedure. On the other hand, these
may be the ones most likely to `detect' small residual
systematics; it seems conservative to distribute the
blame for any small disagreement uniformly.
One might also query whether this correction should be
applied at all $k$; for many models, the disagreement
is worst at high $k$. We shall stick with the simplest
procedure, since the quoted errors are usually much larger
at low $k$.

Of our free parameters, only two are really important:
$\Omega$ and a measure of the overall level of
fluctuations. We take the IRAS bias parameter to
play this latter role. Once these two are specified,
the other bias parameters are well determined --
principally from the data at small $k$, where we
are in the linear regime. The best-fit values
depend only very slightly on the two controlling
parameters, and for all allowed models are close to
\be
b_{\ss A}:b_{\ss R}:b_{\ss O}:b_{\ss I} =
4.5:1.9:1.3:1,
\ee
to within 6 per cent rms.
We now show likelihood plots for the remaining two
parameters, $\Omega$ and $b_{\ss I}$. Contours of
likelihood are displayed in Figure 5, distinguishing
the cases $\Omega_v=0$ (open) and $\Omega_m+\Omega_v=1$ (flat).
Two main features are visible on these plots:
the data appear to demand a significant degree
of redshift-space distortion, with the optimal
model having
\be
{\Omega^{0.6}\over b_{\ss I}}=1.0\pm 0.2
\ee
in both cases (rms error). Models satisfying
this constraint in which both $\Omega$ and $b$
are large are allowed, corresponding to models well
in the linear regime. However, low-bias models
appear to be less favoured: for low $\Omega$,
the best models have $b_{\ss I}\simeq 0.8$.
For the case of flat models, there is a certain
bimodality, with the preferred values of
$b_{\ss I}$ for $\Omega=0.1$ being 0.8 and 0.25.
However, the heavily antibiased branch of
solutions can probably be excluded on other
grounds, and we ignore it hereafter.
At the 90 per cent confidence level,
this analysis requires $\Omega>0.14$.
The various reconstructions of the linear power spectrum
for the case $\Omega=b_{\ss I}=1$
are shown superimposed in Figure 6, and display
an impressive degree of agreement.
This argues very strongly that what we measure
with galaxy clustering has a direct relation to
mass fluctuations, rather than the large-scale
clustering pattern being an optical illusion
caused by non-uniform galaxy-formation
efficiency (Bower et al. 1993). If this were
the case, the shape of spectrum inferred from
clusters should have a very different shape at
large scales, contrary to observation.

The detection of redshift-space distortions
is based largely on the inclusion of the APM survey,
since it is the only real-space measurement used here.
If this dataset is
removed from the analysis, small values of
$\Omega^{0.6}/b_{\ss I}$ are no longer excluded.
An upper limit at $\Omega^{0.6}/b_{\ss I}\ls 2$
can still be set; this comes primarily from the
cross-correlation data. In real space, the cross-correlation
should be the geometric mean of the two auto-correlation
results. Because of the different effects of the
redshift-space mapping, however, this is no
longer true when redshift-space distortions become
large. The observed cross-correlations thus set
a limit to how strong the distortion can be.
Some independent confidence in the
detection of non-zero distortion can be gained from
the work of Saunders, Rowan-Robinson \& Lawrence (1992).
They deduced the real-space correlation function
for IRAS galaxies: $\xi(r)=(r/r_0)^{-\gamma}$, with
$r_0=3.78\pm0.14\,h^{-1}\rm Mpc$ and
$\gamma=1.57\pm0.03$. If we convert this to a power
spectrum, it lies lower than the QDOT results of
FKP by a factor $1.61\pm0.26$
over the range $0.05h<k<0.15h\;\rm Mpc^{-1}$.
This corresponds to $\Omega^{0.6}/b_{\ss I}=0.75\pm0.25$, in
good agreement with the figure deduced above, and
provides independent evidence for the detection
of significant redshift-space distortion. This lower
degree of real-space clustering is also in agreement
with our ratio of 1.3 between optical and IRAS
bias factors. The Saunders et al. figure for
$r_0$ predicts $r_0=5.3\,h^{-1}\rm Mpc$ for optically-selected
galaxies in real space, which is very close to the canonical
value. IRAS galaxies have a slightly smaller value of
$\gamma$, but this only produces an important
change in relative power on scales rather smaller
than those probed here.

The conclusion that models with $\Omega^{0.6}/b\simeq 1$
and low $\Omega$ are not allowed stems from the
effect of nonlinearities: the true level of
mass fluctuations in such models would be very
high. Moreover, going to low $\Omega$ increases
the effect of nonlinearities, as discussed above;
this trend is less marked for the flat models,
which is why low densities are not so
strongly excluded in that case.
It is easy to see how this conclusion arises
by referring to Figure 6. This shows that the
linear power spectra inferred from galaxy and
cluster data agree down to $k\simeq 0.3h\,\rm Mpc^{-1}$,
where $\Delta^2\simeq 1$ in the best-fitting case.
If we assume a higher normalization, the effect
of nonlinearities in this case is to add power, so the linear
reconstruction from galaxy data would become
very flat at high $k$ (cf. Figure 3). However, this would
disagree with the cluster data, which would still
indicate a steep power spectrum, since we have
assumed that the clusters give the linear result directly.
This is a general problem with highly evolved models:
since nonlinearities change the shape of the power
spectrum at $\Delta^2\simeq 1$, and especially so
for low densities, it requires something of a
conspiracy for the nonlinear power spectrum to be a
featureless power law (see Gott \& Rees 1975).
However, on the present assumptions, extreme nonlinear evolution
should steepen the galaxy correlations faster
than those for clusters, and yet they empirically
have much the same slope. The easiest way of
understanding this is to say that the
degree of nonlinearity is only mild.
This is certainly an issue which merits further
investigation, and a detailed simulation
of cluster formation in a
highly nonlinear low-density model would be most valuable.
In the meantime, it is interesting to note that
the constraints we have drawn here on density
and bias are very similar to those obtained
in a completely independent way by the POTENT
group in their analysis of the peculiar-velocity
field (Dekel et al. 1993).

Table 1 gives the final data for the mean reconstructed
power spectrum, for the case $\Omega=b_{\ss I}=1$.
The data have been averaged in bins of
width 0.1 in $\log_{10}$(wavenumber)
and the errors quoted are standard errors.
These numbers are plotted in Figure 7, and will be compared
with models in the next Section; as will be shown there,
the data are consistent with a smooth and featureless
power spectrum, despite the small size of the errors.
One of the pleasant features of our result is that
the power spectrum is only weakly dependent on
model parameters. For $\Omega=1$, the power is
not so sensitive to $b$, because in redshift
space (the majority of the data) we measure
\be
\Delta_z^2\propto b^2
\left(1+{2\over3}[f/b] + {1\over 5}[f/b]^2\right).
\ee
The overall power correction factor thus scales
only as $b^{10/7}$ for $b$ close to unity. This
can be used to rescale our `standard' result to
some other desired value of $b$, given $\Omega=1$.
For low densities, an empirical formula for the
scaling of the linear mass spectrum in the
present analysis is
\be
\Delta^2_{\ss L} \propto  \Omega^{-0.3}.
\ee

It is convenient to be able to compare the results here
with another common measure of the amplitude of linear
mass fluctuations. This is $\sigma_8$: the linear-theory rms
density contrast when averaged over spheres of radius $8h^{-1}$
Mpc:
\be
\sigma_R^2 =\int \Delta^2(k)\;{dk\over k}
\;{9\over (kR)^6}[\sin kR -kR\cos kR]^2.
\ee
The squared window function weighting the power spectrum
is very close to a Gaussian $W_k^2 =\exp[-k^2 R^2/5]$, and
so $\sigma_R^2$ is just $\Delta^2(k)$ at some effective
wavenumber:
\ba
\sigma_R^2 = \A\Delta^2(k_R) \\
k_R = \A\left[{([n+1]/2)\,!\over 2}\right]^{1/(n+3)}
{\sqrt{5}\over R},
\ea
where $n$ is the effective power-law index of the power
spectrum. As before, this approximation is within a few per cent of
the exact integration provided $n\ls 0$.
On the scales of interest, the effective index is close to
$-1.5$ and so the effective wavenumber for $\sigma_8$ is
$k =0.20$. Using the above scalings, we get
\be
\sigma_8 =  0.75\; \Omega^{-0.15},
\ee
with a formal rms uncertainty of 13 per cent.

The significance of $8h^{-1}$ Mpc as a normalization
scale is that $\sigma_8$ is of order unity and thus
its value can be probed by observations of weakly
nonlinear structures such as galaxy clusters. White,
Efstathiou \& Frenk (1993) discuss this constraint,
and deduce $\sigma_8 =0.57\,\Omega^{-0.56}$ for spatially
flat models (although the scaling should be very similar
for open models), to within a tolerance of roughly $\pm 10$
per cent. The precise meaning of their uncertainty is
hard to quantify, but it seems intended to give hard
limits, rather than an rms.
The agreement with our results is very good;
the $\Omega$ dependence is steeper, but the
disagreement in $\sigma_8$ is only a factor 1.4
even for $\Omega=0.2$.

\section{POWER-SPECTRUM DATA AND MODELS}

\subsection{CDM-like models}

It is interesting to ask if the power spectrum
contains any features, or whether it is consistent
with a single smooth curve. In fact, a variety of
simple models describe the data from Table 1
very well within the errors. Consider the
fitting formula used by Peacock (1991), which
is just a break between two power laws:
\be
\Delta^2(k)={(k/k_0)^\alpha\over 1+(k/k_c)^{\alpha-\beta}}.
\ee
This works well, with
\ba
k_0 \A = 0.29\pm0.01\; h\, {\rm Mpc}^{-1} \\
k_1 \A = 0.039\pm0.002\; h\, {\rm Mpc}^{-1} \\
\alpha \A = 1.50\pm0.03 \\
\beta \A = 4.0\pm0.5.
\ea
A value of $\beta=4$ corresponds to a scale-invariant
spectrum at large wavelengths.

An more physical alternative is the CDM power
spectrum, which is $\Delta^2(k)\propto k^{n+3}T_k^2$.
We shall use the BBKS approximation for the transfer
function:
\ba
T_k= \A {\ln(1+2.34q)\over 2.34q} \;\times \nn
     \A [1+3.89q+(14.1q)^2 +(5.46q)^3 +(6.71q)^4]^{-1/4},
\ea
where $q\equiv k/[\Omega h^2\; {\rm Mpc}^{-1}]$. Since observable
wavenumbers are in units of $h\,{\rm Mpc}^{-1}$, the shape parameter
is the apparent value of $\Omega h$.
This scaling applies for models with zero baryon content,
but there is an empirical scaling that can account
for the effect of baryons, and which deserves to
be more widely known.
Figure 8 shows a compilation of CDM transfer
functions taken from Holtzman (1989).
When plotted against $k/\Omega h^2$, there is
a strong dependence on baryon density: high
baryon content mimics low CDM density. If we instead
use the scaling
\be
T_k(k)=T_{\ss BBKS}(k/[\Omega h^2\,\exp -2\Omega_{\ss B}]),
\ee
then all the curves lie on top of one another
to a few per cent tolerance. We shall henceforth
use the term `$\Omega h$' to refer to the BBKS
fitting parameter, on the understanding that it
means the combination $\Omega h \exp -2\Omega_{\ss B}$.
Our results will hence differ slightly from those
of Efstathiou, Bond \& White (1992), who defined a
parameter $\Gamma$ which is almost $\Omega h$.
Unfortunately, they scaled to a `standard' CDM
model with $\Omega_{\ss B}=0.03$, with the result
that $\Gamma=1.06\Omega h$.

Fitting the CDM model to our data also results
in a satisfactory $\chi^2$ and requires the
parameters
\be
\Omega h =  0.255\pm0.017 + 0.32({1/n}-1),
\ee
in agreement with many previous arguments suggesting
that a low-density model is needed.
The fit of this and other models is illustrated in Figure 9.
For any reasonable values of $h$ and baryon density,
a high-density CDM model is not viable. Even a high degree
of `tilt' in the primordial spectrum (Cen et al. 1992)
does not help reach the required $\Omega h\simeq 0.75$.
The alternatives are either to retain the CDM model,
but assume that some piece of unknown physics has
produced a transfer function which looks like
a low-density model, or to adopt a low density,
or to go for something else entirely.
As far as low densities are concerned, note that the
popular choice of $\Omega=0.2$
(e.g. Kauffmann \& White 1992) will overshoot
and yield too low values of $\Omega h$.
More viable alternatives with high density are
either mixed dark matter (Holtzman 1989; van Dalen
\& Schaefer 1992; Taylor \& Rowan-Robinson 1992;
Davis, Summers \& Schlegel 1992; Klypin et al. 1993;
Pogosyan \& Starobinsky 1993),
or non-Gaussian pictures such as cosmic strings + HDM,
where the lack of a detailed prediction for
the power spectrum helps ensure that the model is
not yet excluded (Albrecht \& Stebbins 1992).
Mixed dark matter seems rather ad hoc, but may
be less so if it is possible to produce both
hot and cold components from a single particle,
with a Bose condensate playing the role of
the cold component (Madsen 1992; Kaiser, Malaney
\& Starkman 1993).
However, the shape of the MDM spectrum is not
really all that close to the spectrum deduced here: it bends
much more sharply, and is very flat on small
scales. At the quasilinear scale $k=0.2h\;{\rm Mpc}^{-1}$,
the local power-law index for the MDM model is
about $n=-2.2$, as opposed to our empirical
value $n\simeq -1.5$.
If the good fit of a low-density CDM transfer function is taken
literally, then perhaps this is a hint that the
epoch of matter-radiation equality needs to be delayed.
An approximate doubling of the number of relativistic degrees of
freedom would suffice -- but  this would
do undesirable violence to primordial
nucleosynthesis: any such boost would have
to be provided by a particle which decays after nucleosynthesis.
The apparent value of $\Omega h$ depends on the mass
and lifetime of the particle roughly as
\be
\left.\Omega h\right|_{\rm apparent} =
\Omega h\; [1+(m_{\rm keV} \tau_{\rm years})^{2/3}]^{-1/2}
\ee
(Bardeen, Bond \& Efstathiou 1987; Bond \& Efstathiou 1991),
so a range of masses is possible.
Apart from making the observed large-scale structure,
such a model yields a small-scale enhancement
of power which could lead to early galaxy formation.
Whether the required particle physics is at all
plausible remains to be seen, but the model is
arguably the most attractive of those currently available.

An important general lesson to be drawn from this section is the
lack of large-amplitude features in the power spectrum.
This is a strong indication that collisionless
matter is deeply implicated in forming large-scale
structure. Purely baryonic models contain
large bumps in the power spectrum around
the Jeans' length prior to recombination
($k\sim 0.03\Omega h^2\;\rm Mpc^{-1}$), whether
the initial conditions are isocurvature or
adiabatic (e.g. Section 25 of Peebles 1993).
It is hard to see how such features can
be reconciled with the data.

\subsection{Peculiar velocities}

The mass power spectrum has a direct application in
predicting the cosmological peculiar velocity field.
The 3D rms velocity for clumps averaged over
some window is
\be
\sigma_v^2=H^2 \,f(\Omega)^2\int \Delta^2(k)\; {dk\over k^3}\; W_k^2,
\ee
so we can use the power spectrum to make a direct
prediction of this quantity, which is shown in
Figure 10 for the case of spheres of varying radii.
The velocity power spectrum ($\propto k^{-2}\Delta^2(k)$)
peaks around the break in the power spectrum at $k\simeq 0.03h\,
\rm Mpc^{-1}$, and so the predicted velocities
decline rapidly for spheres which filter out this
scale.

For $\Omega=1$, the predicted velocities are very
reasonable. If we model the local group as a sphere
of radius $5h^{-1}$ Mpc, the 3D rms is 680 $\rm km\,s^{-1}$,
as against the observed 600 one-point local measurement
(the answer is very insensitive to the size used
to define the local group).
Figure 10 also shows
the deduced velocities from the POTENT group
(Bertschinger et al. 1990) for
spheres of radius 40 and 60 $h^{-1}$Mpc, which also
agree well. However, the predictions are completely
inconsistent with the velocity of 842 $\rm km\,s^{-1}$
for the local sphere out to $150h^{-1}$Mpc claimed
by Lauer \& Postman (1993). The predicted 3D rms
for this scale is only 140 $\rm km\, s^{-1}$. Even
if we allow that their weighting scheme might
reduce the effective radius of their sphere
(they weight each radial shell equally), there
remains a qualitative discrepancy. If this
results were to be confirmed, it would probably
indicate a large feature in the power spectrum on
scales beyond those probed here ($k\ls0.01h\,\rm Mpc^{-1}$).

The empirical power spectrum deduced here thus seems
to agree extremely well with large-scale velocity data.
The crucial test for $\Omega=1$ models, however, has
often been the small-scale velocity dispersion.
The preferred low-$\Omega h$ model predicts a pairwise dispersion
at $1h^{-1}$ Mpc separation of about $\sigma_\parallel=550\;
\rm km\,s^{-1}$ (Mann 1993), which is interestingly
close to more recent observational data (Mo, Jing \& B\"orner 1993).

\subsection{CMB anisotropies}

We now relate the measurement of mass fluctuations
on scales of several hundred Mpc to those implied
on larger scales from the measurement of CMB
fluctuations by the COBE team (Smoot et al. 1992).
This is a subject which has advanced rapidly since the
original detection, with a more widespread appreciation
of the possible contribution of gravitational
waves to the anisotropy (following the original
insight of Starobinsky 1985). We therefore
distinguish explicitly between scalar and tensor
contributions to the CMB fluctuations by using appropriate
subscripts. The former
category are those described by the Sachs-Wolfe effect,
and are gravitational potential fluctuations
that relate directly to mass fluctuations. For a
Gaussian beam of FWHM $2.35\sigma$, the correlation
function of the microwave sky is
\be
C_{\ss S}(\theta) ={1\over 4\pi}\sum_\ell (2\ell+1)\,
 W_\ell^2\, C_\ell\, P_\ell(\cos\theta),
\ee
where $P_\ell$ are Legendre polynomials, and $W_\ell =\exp(-\ell^2\sigma^2/2)$.
The coefficients $C_\ell$ are
\be
C_\ell =16\pi\, {\Omega^2\over g^2(\Omega)}
 \int(k[2c/H_0])^{-4}\Delta^2(k)\;j_\ell^2(k R_{\ss H})\;
{dk\over k}
\ee
where $j_\ell$ are spherical Bessel functions (see Peebles 1982).
The length $R_{\ss H}$ is the present comoving horizon size
\ba
R_{\ss H} = \A {2c\over \Omega H_0}\quad\rm (Open) \\
          \simeq \A {2c\over \Omega^{0.4} H_0}\quad\rm (Flat)
\ea
(Vittorio \& Silk 1991)
and the function $g(\Omega)$ is the linear growth
suppression factor relative to $\Omega =1$, as discussed earlier.
These formulae strictly apply only to spatially flat
models, since the notion of a scale-free
spectrum is imprecise in an open model. Nevertheless,
since the curvature radius subtends an angle of
$\Omega/[2(1-\Omega)^{1/2}]$, normalizing to COBE in
an open model should not be a very bad approximation
until we reach $\Omega\ls 0.2$. We shall therefore
ignore this uncertainty in what follows.

In the case of the COBE measurements, the simplest and
most robust datum is just the sky variance convolved
to $10^\circ$ FWHM, i.e. $C_{\ss S}(0)$ in the above
expression with $\sigma =4.25^\circ$.
This can be converted into an integral
over the power spectrum multiplied by a window function
which is a sum over Bessel functions. In practice,
it is convenient to have a simpler expression for
the window, and it turns out that this can be achieved
to almost perfect accuracy by using a
small-angle approximation:
\ba
C_{\ss S}(0) = \A {\Omega^2\over g^2(\Omega)}\, \int
4(k[2c/H_0])^{-4}\Delta^2(k)\;W^2(k R_{\ss H})\;
{dk\over k} \\
W^2(y) = \A [1-j_0^2(y)-3j_1^2(y)]\,F(y\sigma)/(y\sigma),
\ea
where $F(x)$ is Dawson's integral. The terms involving
Bessel functions correspond to the subtraction of monopole
and dipole terms.
The window function is relatively sharply peaked
and so the COBE variance essentially picks out
the power at a given scale. For the case of
$\sigma =0.0742$ (FWHM of $10^\circ$), the result
is very well fit by
\ba
C_{\ss S}(0) = \A 1.665\, {\Omega^2\over g^2(\Omega)}
\, [4(k_{\ss S}[2c/H_0])^{-4}
\Delta^2(k_{\ss S})] \\
k_{\ss S}R_{\ss H} = \A 7.29 + 2.19(n-1)
\ea
The observed value is $C^{1/2}(0) =1.10\pm0.18\times 10^{-5}$
(Smoot et al. 1992). For scale-invariant spectra, this
corresponds to an rms quadropole of $Q_{\rm rms} =15.0\pm2.5\;\mu$K.
For $\Omega =1$, this translates to a normalization of
$\epsilon =2.6\pm0.4\times 10^{-5}$ in the notation of
Peacock (1991).
How well does this amplitude match on to the clustering
observed at 100 Mpc wavelengths? If we stick to asymptotically
scale-invariant spectra, the agreement is very good. The
CDM fit shown in Figure 9 requires
\be
\epsilon =3.25\pm0.18\times 10^{-5}.
\ee
This is slightly higher than the COBE measurement, but
well within experimental error. If the large-scale normalization
is forced to be $\epsilon=2.6\times 10^{-5}$, the
best-fitting CDM shape changes to $\Omega h=0.31$.

For a more general comparison, it is convenient to
define a reference datum at the largest scale where
our data are still accurate. From Table 1, we take
this to be $\Delta^2(k=0.028h)= 0.0087 \pm 0.0023$.
At this point, there is still some
curvature in the power spectrum: the $\Omega h=0.25$
transfer function is $T_k=0.61$ and the effective
transfer function defined by the two power-law
formula is $T_k=0.80$. We shall adopt a compromise
$T_k$=0.70 and hence deduce
\be
\Delta^2(k=0.028h)= 0.018 \pm 0.0023
\ee
as our best estimate of the true level of
any primordial power-law fluctuations on these
scales (subject to scalings as above if $\Omega\ne 1$).
We can now use our earlier discussion of the
COBE data to predict this small-scale fluctuation,
ignoring for the moment any gravity-wave
contribution. The answer is
\ba
\Delta^2(k=0.028h) = \A 0.014 \, \exp[3.2(n-1)]\, \Omega^{-0.7}\quad \rm (open)
\\
                   = \A 0.014 \, \exp[3.2(n-1)]\, \Omega^{-1.6}\quad \rm (flat)
{}.
\ea
Thus, if we adopt $\Omega=1$, there is a very good agreement with
scale-invariance: $n=1.08\pm0.04$. Conversely, tilted models
do not match large and small scales very well: for $n=0.7$, the
predicted power near 100 Mpc is too small by a factor 3.
Things get worse if gravity waves are included: a prediction of
many inflationary models is that
\be
{C_\ell^{\ss G}\over C_\ell^{\ss S}} \simeq 6(1-n)
\ee
(e.g. Liddle \& Lyth 1991; Lidsey \& Coles 1991;
Lucchin, Matarrese \& Mollerach 1992; Souradeep \& Sahni 1992),
which decreases the predicted small-scale power by
a further factor 2.8 for $n=0.7$, making a total mismatch
of a factor 8. It is inconceivable that our analysis of
the 100-Mpc scale power could be in error by this amount.
Thus, although tilted models may be attractive in
removing the 1-degree `bump' in the predicted microwave
sky (Crittenden et al. 1993) and allowing
consistency with intermediate-scale CMB
experiments, it seems implausible that this can be
the correct solution, at least if $\Omega=1$.
To allow a tilted model with $n=0.7$, we need
$\Omega\simeq 0.06$ or 0.3 respectively in the open and
flat cases.

\section{SUMMARY}

We have analysed a compilation of recent measures
of galaxy clustering, under the assumption of
underlying Gaussian mass fluctuations.
We have presented new methods for dealing
analytically with the modifying effects
of nonlinear evolution and redshift-space distortions,
and their effect on the power spectrum.
Applying these methods to the data leads
to a consistent determination of the linear
mass spectrum, with the following properties.

\japitem{(i)}The relative bias factors
for Abell clusters, radio galaxies, optical
galaxies and IRAS galaxies must be in the
ratios $b_{\ss A}:b_{\ss R}:b_{\ss O}:b_{\ss I} =
4.5:1.9:1.3:1$, to within 6 per cent rms.

\japitem{(ii)}The data require a significant degree
of redshift-space distortion: $\Omega^{0.6}/b_{\ss I}
 = 1.0 \pm 0.2$.

\japitem{(iii)}Low values of $\Omega$ and bias
are disfavoured because nonlinear evolution would
spoil the agreement in shape between galaxy and
cluster power spectra. Both this and the previous
conclusion are in good agreement with
independent studies based on peculiar velocity fields.

\japitem{(iv)}The linear power spectrum is smooth
and featureless, and is well described by a zero-baryon
CDM model with $\Omega h=0.25$.

\japitem{(v)}The amplitude of 100-Mpc power matches
well onto that inferred from COBE provided the
primordial spectrum was close to scale-invariant.
Tilted models that postulate a dominant gravity-wave
CMB component are difficult to reconcile with our data.

\section*{ACKNOWLEDGEMENTS}

S.J.D. is supported by a SERC research studentship.
We thank Hugh Couchman for the use of his AP$^3$M code,
Carlton Baugh for communicating the APM
power-spectrum data, and Guinevere Kauffmann \&
Simon White for providing $N$-body data.

\section*{REFERENCES}

\ref Albrecht A., Stebbins A., 1992, \prl, 69, 2615
\ref Babul A., Postman M., 1990, \apj, 359, 280
\ref Bardeen J.M., Bond J.R., Kaiser N., Szalay A.S., 1986, \apj, 304, 15
(BBKS)
\ref Bardeen J.M., Bond J.R., Efstathiou G., 1987, \apj, 321, 28
\ref Baugh C.M., Efstathiou G., 1993, \mn, 265, 145
\ref Bertschinger E., Dekel A., Faber S.M., Dressler A., Burstein D., 1990,
\apj, 364, 370
\ref Bond J.R., Couchman H.M.P., 1988, in Coley A., Dyer C.C., Tupper B.O.J.,
eds, Proc.  Second Canadian Conference on General Relativity \& Relativistic
Astrophysics (World Scientific)
\ref Bond J.R., Efstathiou G., 1991, Phys. Lett. B, 265, 245
\ref Bouchet F.R., Strauss M.A., Davis M., Fisher K.B., Yahil A., Huchra J.P.,
1993, \apj, 417, 36
\ref Bower R.G., Coles P., Frenk C.S., White S.D.M., 1993, \apj, 405, 403
\ref Broadhurst, T.J., Ellis, R.S., Koo, D.C.,  Szalay, A.S., 1990, Nature, 343
726
\ref Carroll S.M., Press W.H., Turner E.L., 1992, \annrev, 30, 499
\ref Cen R., Gnedin N.Y., Kofman L.A.,  Ostriker J.P., 1992, \apj, Lett, 399,
L11.
\ref Cen R., Ostriker J.P., 1992, \apj, 399, L113
\ref Coles P., Plionis M., 1991, \mn, 250, 75
\ref Couchman H.M.P., 1991, \apj, 368, L23
\ref Crittenden R., Bond J.R., Davis, R.L., Efstathiou, G., Steinhardt P.J.,
1993, \prl, 71, 324.
\ref Davis M., Efstathiou G., Frenk C.S., White S.D.M., 1985, \apj, 292, 371
\ref Davis M., Summers F.J., Schlegel D., 1992, \nat, 359, 393
\ref Dekel A., Bertschinger E., Yahil A., Strauss M.A., Davis M., Huchra J.P.,
1993, \apj, 412, 1.
\ref Einasto J., Joeveer M., Saar E., 1980, \mn, 193, 353
\ref Efstathiou G., Bond J.R., White S.D.M., 1992, \mn, 258, 1P
\ref Feldman H.A., Kaiser N.,  Peacock J.A., 1993, \apj, in press (FKP)
\ref Fisher K.B., Davis M., Strauss M.A., Yahil A.,  Huchra J.P., 1993, \apj,
402, 42
\ref Gazta\~naga E., 1992, \apj, 398, L17
\ref Gooding A.K., Park C., Spergel D.N., Turok N., Gott J.R. III, 1992, \apj,
393, 42
\ref Gott J.R. III, Rees M.J., 1975, \astast, 45, 365
\ref Gramann M., Cen R., Bahcall N., 1993, \apj, in press
\ref Hamilton A.J.S., Gott J.R. III, Weinberg D.H., 1986. \apj, 309, 1
\ref Hamilton A.J.S., Kumar P., Lu E.,  Matthews A., 1191, \apj, 374, L1 (HKLM)
\ref Holtzman J.A., 1989, \apjs, 71, 1
\ref Kaiser N., 1984, \apj,  284, L9
\ref Kaiser N., 1987, \mn, 227, 1
\ref Kaiser N., Peacock J., 1991, \apj, 379, 482
\ref Kaiser N., 1987, \mn, 227, 1
\ref Kaiser N., Malaney R.A., Starkman G.D., 1993, \prl, in press
\ref Kauffmann G., White S.D.M., 1992, \mn, 258, 511
\ref Klypin A., Holtzman J., Primak J., Reg\H os E., 1993, \apj, 416, 1
\ref Kofman L., Gnedin N., Bahcall N., 1993, \apj, 413, 1
\ref Kofman L., Gelb J., Bertschinger E., Nusser A., Dekel A., 1993, \apj, in
press.
\ref Lahav O., Lilje P.B., Primak J.R., Rees M.J., 1991, \mn, 251, 128
\ref Lauer T.R., Postman M., 1993. Proc. Milan meeting on observational
cosmology
\ref Liddle A.R., Lyth D., 1992,  Phys Lett, 291B, 391.
\ref Lidsey J.E., Coles P., 1992,  \mn, 258, 57P
\ref Loveday J., Efstathiou G., Peterson B.A., Maddox S.J., 1992, \apj, 400,
L43
\ref Lucchin F., Matarrese S., Mollerach S., 1992, \apj, 401, 49
\ref Madsen J., 1992, \prl, 69, 571
\ref Mann R.G., 1993, PhD thesis, University of Edinburgh
\ref Mann R.G., Heavens A.F, Peacock J.A.,  1993, \mn, 263, 798
\ref Mo H.J., Peacock J.A.,  Xia X.Y., 1993, \mn, 260, 121
\ref Mo H.J., Jing Y.P., B\"orner G., 1993, \mn, 264, 825
\ref Moore B. et al., 1992, \mn, 256, 477
\ref Nusser A., Dekel A., 1993, \apj, 405, 437
\ref Peacock J.A., 1991, \mn,  253, 1P
\ref Peacock J.A., 1992, New insights into the Universe, proc. Valencia summer
school, eds  Martinez V., Portilla M., S\'aez D. (Springer), p1
\ref Peacock J.A., Heavens A.F., 1985, \mn, 217, 805
\ref Peacock J.A., Nicholson D., 1991, \mn, 253, 307
\ref Peacock J.A., West M.J., 1992, \mn, 259, 494
\ref Peebles P.J.E., 1980, The Large--Scale Structure of the Universe.
Princeton Univ. Press, Princeton, NJ
\ref Peebles P.J.E., 1993, Principles of physical cosmology.  Princeton Univ.
Press, Princeton, NJ
\ref Peebles P.J.E., 1973, \apj, 185, 413
\ref Peebles P.J.E., 1982, \apj, 263, L1
\ref Pogosyan D.Yu., Starobinsky A.A., 1993, \mn, 265, 507
\ref Scherrer R.J., 1992, \apj, 390, 330
\ref Smoot G.F. et al., 1992, \apj, 396, L1
\ref Souradeep T., Sahni V., 1992, Mod. Phys. Lett., 7, 3541
\ref Starobinsky A.A., 1985, Sov. Astr. Lett., 11, 133
\ref Strauss M.A., Davis M., Yahil Y., Huchra J.P., 1992, \apj, 385, 421
\ref Saunders W., Rowan-Robinson M., Lawrence A., 1992, \mn, 258, 134
\ref Szalay A.S., Ellis R.S., Koo D.C., Broadhurst T., 1991, After the first
three minutes, proc AIP conference no 222, eds Holt S.S., Bennett C.L., Trimble
V., p261
\ref Taylor A.N., Rowan-Robinson M., 1992,  \nat, 359, 396
\ref van Dalen A., Schaefer R.K., 1992, \apj, 398, 33
\ref Vogeley M.S., Park C., Geller M.,  Huchra, J.P., 1992, \apj, 391, L5
\ref Vittorio N., Silk, J., 1991, \apj, 385, L9
\ref Weinberg D.H., Cole S., 1992, \mn, 259, 652
\ref White S.D.M., Efstathiou G., Frenk C.S., 1993, \mn, 262, 1023
\ref Zeldovich Ya.B., 1970, \astast, 5, 84

\bigskip
\vbox{
\noindent
{\bf Table 1}\quad The linear power-spectrum data, assuming
$\Omega=b_{\ss I}=1$. To scale the data to other values of these
parameters, see Section 5.2.

\vskip 2 truecm

\tabskip 1.2em
\halign{\hfill#\hfill&#\hfill&#\hfill\cr
$k/h\, {\rm Mpc}^{-1}$ & \hfill$\Delta^2(k)$ & \hfill $\pm$ \cr
\cr
  0.014 &  0.0010 &  0.0003 \cr
  0.018 &  0.0013 &  0.0008 \cr
  0.022 &  0.0032 &  0.0009 \cr
  0.028 &  0.0087 &  0.0023 \cr
  0.035 &  0.0196 &  0.0037 \cr
  0.045 &  0.0312 &  0.004  \cr
  0.056 &  0.052  &  0.008  \cr
  0.071 &  0.107  &  0.011  \cr
  0.089 &  0.146  &  0.017  \cr
  0.112 &  0.211  &  0.027  \cr
  0.141 &  0.33   &  0.033  \cr
  0.178 &  0.43   &  0.051  \cr
  0.224 &  0.73   &  0.095  \cr
  0.282 &  1.14   &  0.13   \cr
  0.355 &  1.63   &  0.27   \cr
  0.447 &  1.61   &  0.41   \cr
}

}

\vfill\eject

\section*{FIGURE CAPTIONS}

\noindent{\bf Figure 1}\quad
The nonlinear evolution of power spectra
according to the HKLM method, and the approximate
direct alternative presented here.
The dashed lines show the input linear power
spectra, the solid lines show the
result of numerically integrating the
HKLM method and the one-step solution is shown dotted.
Panel (a) shows a variety of COBE-normalized
CDM models (with $\Omega h = 0.2$, 0.3, 0.4, 0.5;
while (b) shows $\Omega h$=0.5
CDM filtered with different Gaussian windows
($R_f=0.25$, 0.5, 0.75, 1 $h^{-1}$ Mpc), approximating
the effect of warm dark matter.
As expected, our method fails at very high $k$,
where the linear CDM spectra become very flat and
the linear WDM spectra cut off,
but is otherwise excellent.
Note that the effect of the WDM cutoff is
only felt at very large $k$: WDM is not the
explanation for the shape of the power spectrum
around $k=0.1h$.

\noindent{\bf Figure 2}\quad
The generalization of the HKLM function
relating nonlinear power to linear power
(or $\bar\xi$), as in the original method.
The lowest curve is the original HKLM
function for $\Omega=1$; low densities
give a greater nonlinear response.
We show the fitting formula for open models
only, but what matters in general is
just the $\Omega$-dependent linear growth
suppression factor.

\noindent{\bf Figure 3}\quad
The inverse of the generalized HKLM procedure, as
applied to a power-law power spectrum
$\Delta^2(k)=(k/k_c)^{1.8}$ (in correlation-function
terms, $r_0=0.945/k_c$). Open models with
$\Omega=1$, 0.5, 0.3, 0.2, 0.1 are considered.
Note that the effects of nonlinearities in this
case are rather small for high densities, but
for low densities the required linear initial
conditions are very flat for $k\gs k_c$.
This flat case is one where our
approximate inversion of the HKLM procedure
will not be perfect (cf. Figure 1); nevertheless,
any errors will be small in comparison with
the systematic feature in the power spectrum
required around $\Delta^2\sim 1$.
It clearly requires something of a conspiracy
to achieve a scale-free nonlinear spectrum
in a low-density model.

\noindent{\bf Figure 4}\quad
The raw power-spectrum data used in this analysis.
All data with the exception of the APM power
spectrum are in redshift space. The two
lines shown for reference are the transforms of the
canonical real-space correlation functions
for optical and IRAS galaxies ($r_0=5$ and $3.78
\;h^{-1}\,\rm Mpc$ and slopes of 1.8 and 1.57
respectively).

\noindent{\bf Figure 5}\quad
Contours of relative likelihood based on
the degree of agreement of the various estimates
of linear power spectra.
At each ($\Omega$,$b_{\ss I}$) point, the other
bias factors have been optimized.
We distinguish
the cases $\Omega_v=0$ (open) and $\Omega_m+\Omega_v=1$ (flat).
Contours are plotted at what would be the 50, 90,
95, 99, 99.5 per cent confidence levels in
a two-dimensional Gaussian (i.e. $\Delta\ln {\cal L}
= 0.69$, 2.3, 3.0, 4.6, 5.3).

\noindent{\bf Figure 6}\quad
The power-spectrum data from Figure 4, individually
linearized assuming $\Omega=b_{\ss I}=1$. There is
an excellent degree of agreement, particularly
in the detection of a break around $k=0.03h$.

\noindent{\bf Figure 7}\quad
The linearized data of Figure 6, averaged over
bins of width 0.1 in $\log_{10}k$.
This plot assumes $\Omega=b_{\ss I}=1$; for
lower densities the power increases
slightly, as described in the text.

\noindent{\bf Figure 8}\quad
A set of CDM transfer functions, using the
fitting formulae of BBKS for zero baryon content,
and those of Holtzman (1989) for $\Omega_{\ss B}=0.01$,
0.03, 0.05, 0.1. Models with $h=1$ are
shown as solid lines, $h=0.5$ are dotted. When
plotted (a) against $k/\Omega h^2$, the varying
baryon content causes variations in the
curves, with higher $\Omega_{\ss B}$ producing
greater suppression of power. However,
this can be scaled away (b) by plotting
against the combination $k/[\Omega h^2\exp -2\Omega_{\ss B}]$.
In these terms, the CDM transfer function has
a universal shape, which can be described by the
zero-baryon formula of BBKS.
The scaling becomes noticeably imperfect for $\Omega_{\ss B}
\gs 0.3$, but is very good for the models plotted
here.

\noindent{\bf Figure 9}\quad
The averaged linear power-spectrum data of
Figure 7, compared to various CDM models.
These assume scale-invariant initial conditions,
with the same large-wavelength normalization.
Different values of the fitting
parameter $\Omega h=0.5$, 0.45, \dots 0.25, 0.2
are shown. The best-fit model has
$\Omega h=0.25$ and a normalization which is
$1.5\sigma$ higher than COBE if $\Omega=1$
and gravity-wave anisotropies are negligible
($\epsilon=3.25\times 10^{-5}$).

\noindent{\bf Figure 10}\quad
The predicted 3D rms velocity of spheres as
a function of radius, assuming $\Omega=1$. This is
based on the two power-law fitting formula for
the power spectrum. For low densities the velocities
are reduced, but by less than the normal $\Omega^{0.6}$
factor, because the inferred mass fluctuations rise
in that case (see text). The plotted points are
the local group motion assigned to a radius of
$5h^{-1}$Mpc, and motion of larger spheres
taken from the POTENT group (Bertschinger et al. 1990).
Although the local-group motion is very well determined,
it is assigned a fractional error of $1/\sqrt{6}$ to allow
for fluctuations in the 3D rms velocity seen by different observers.

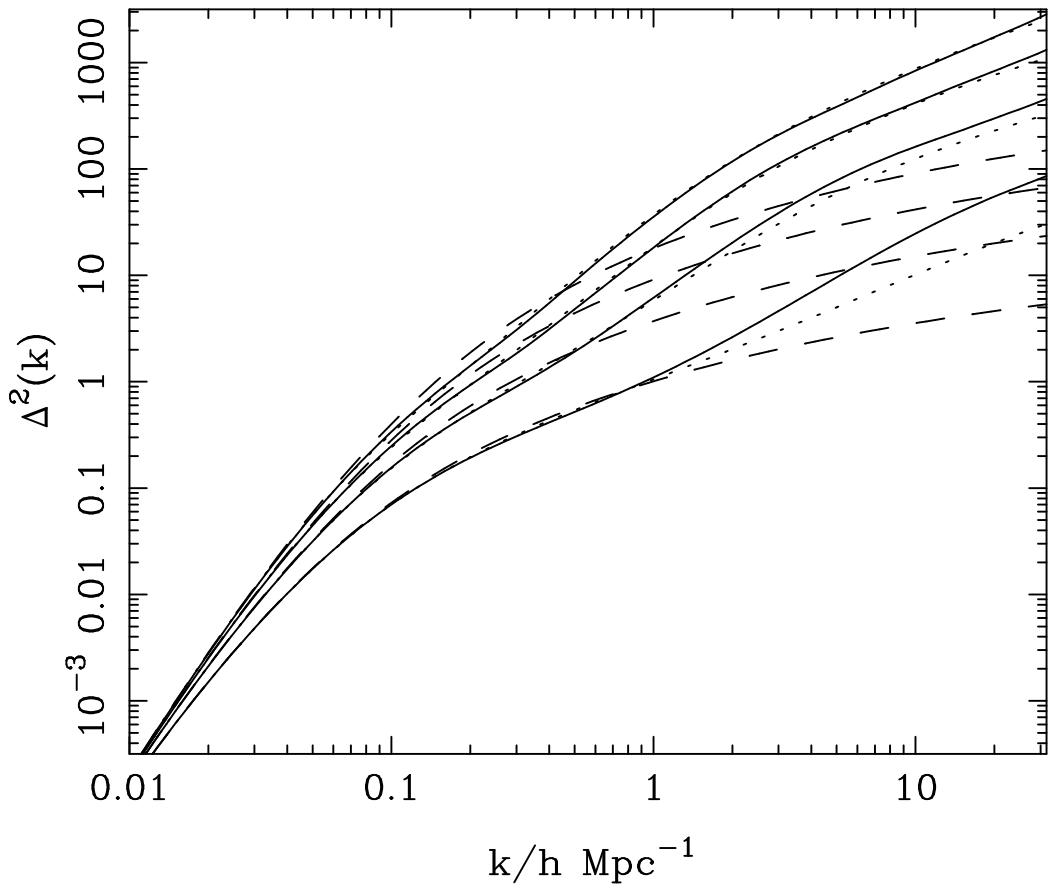
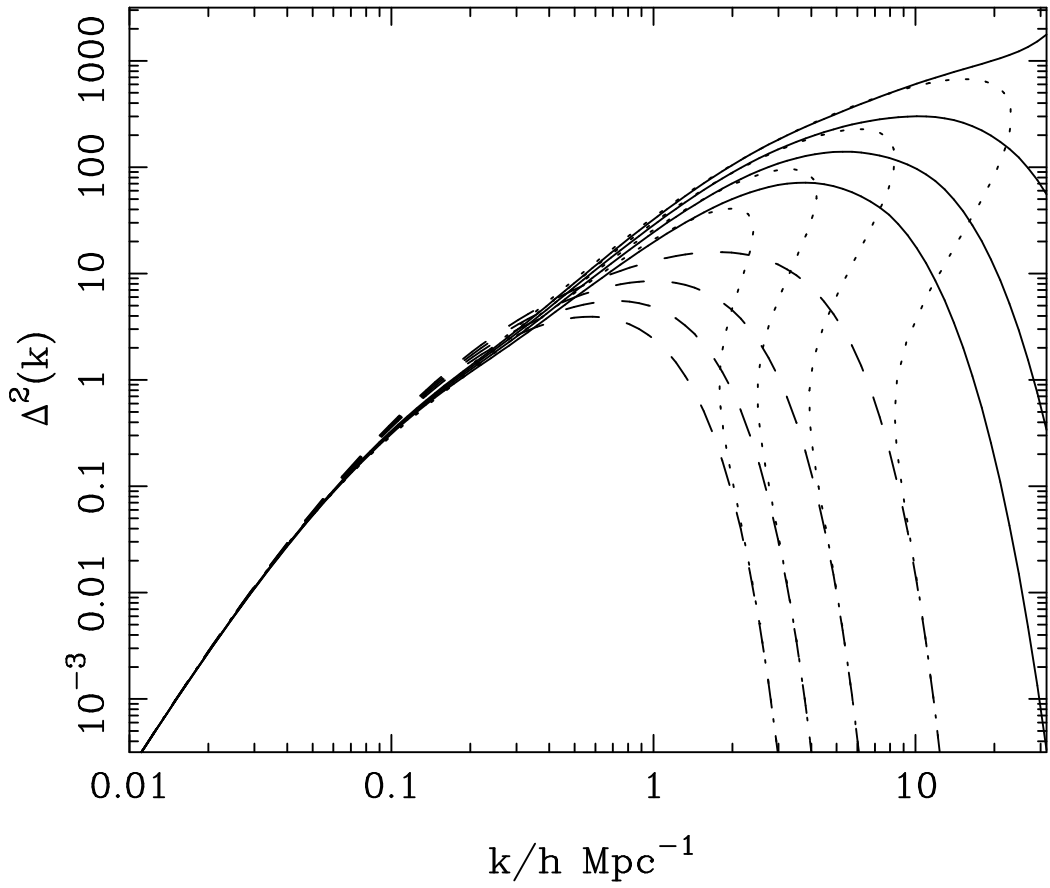
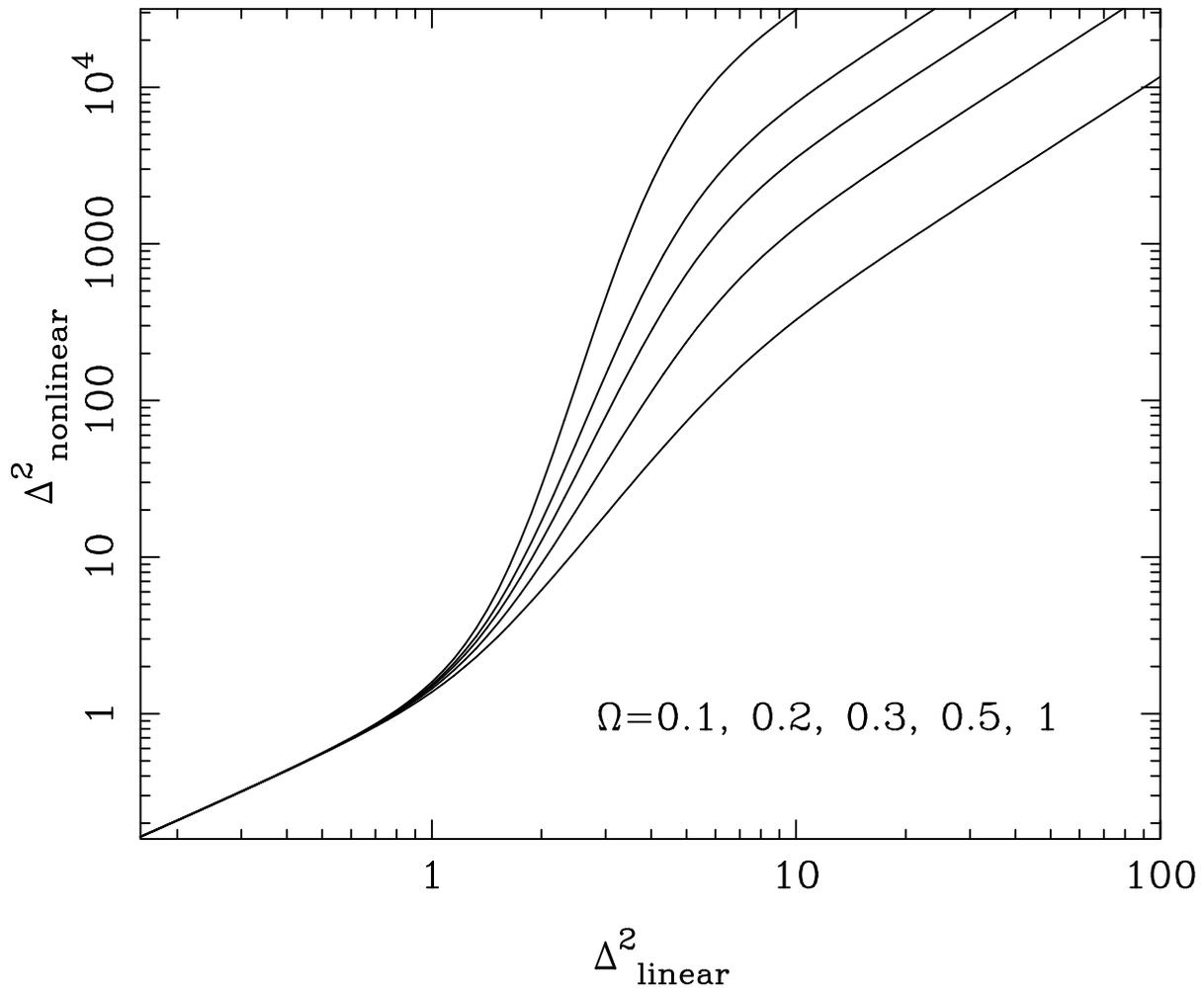
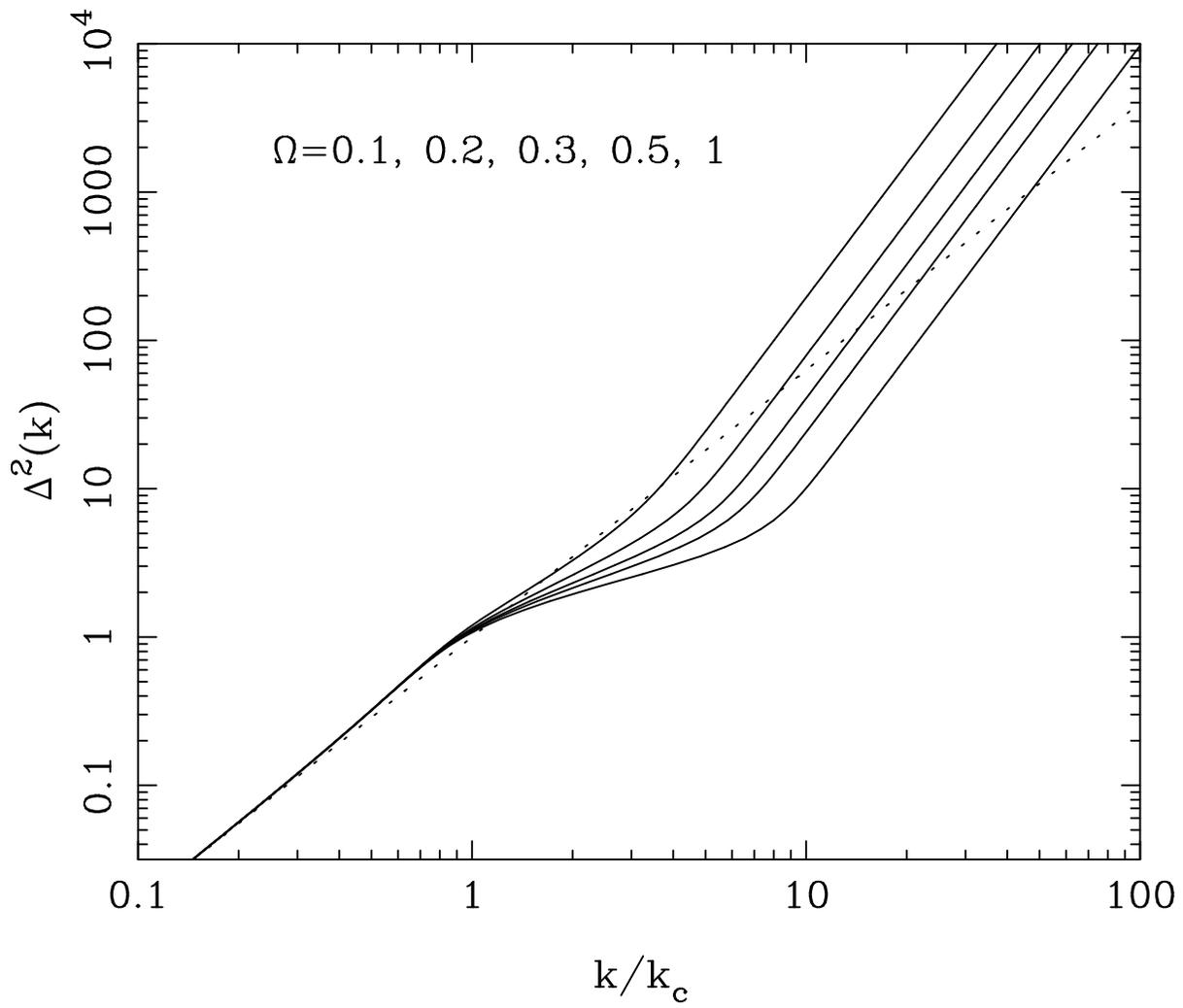
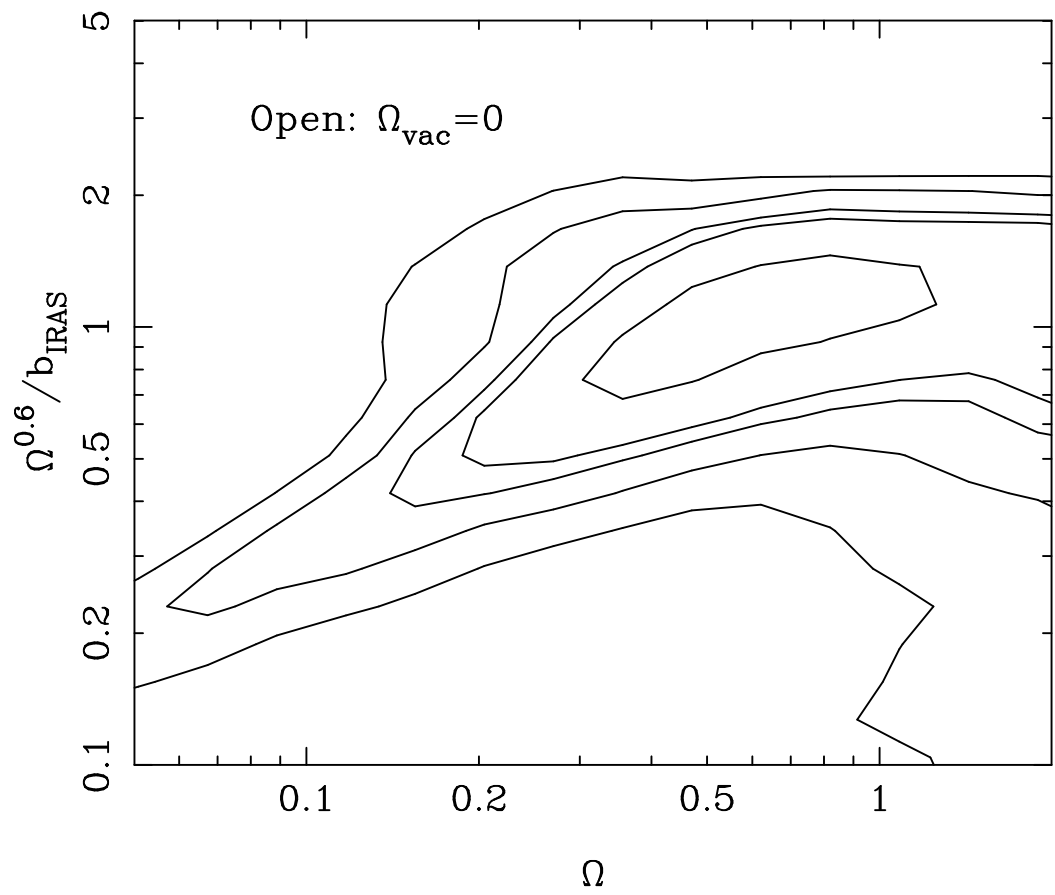
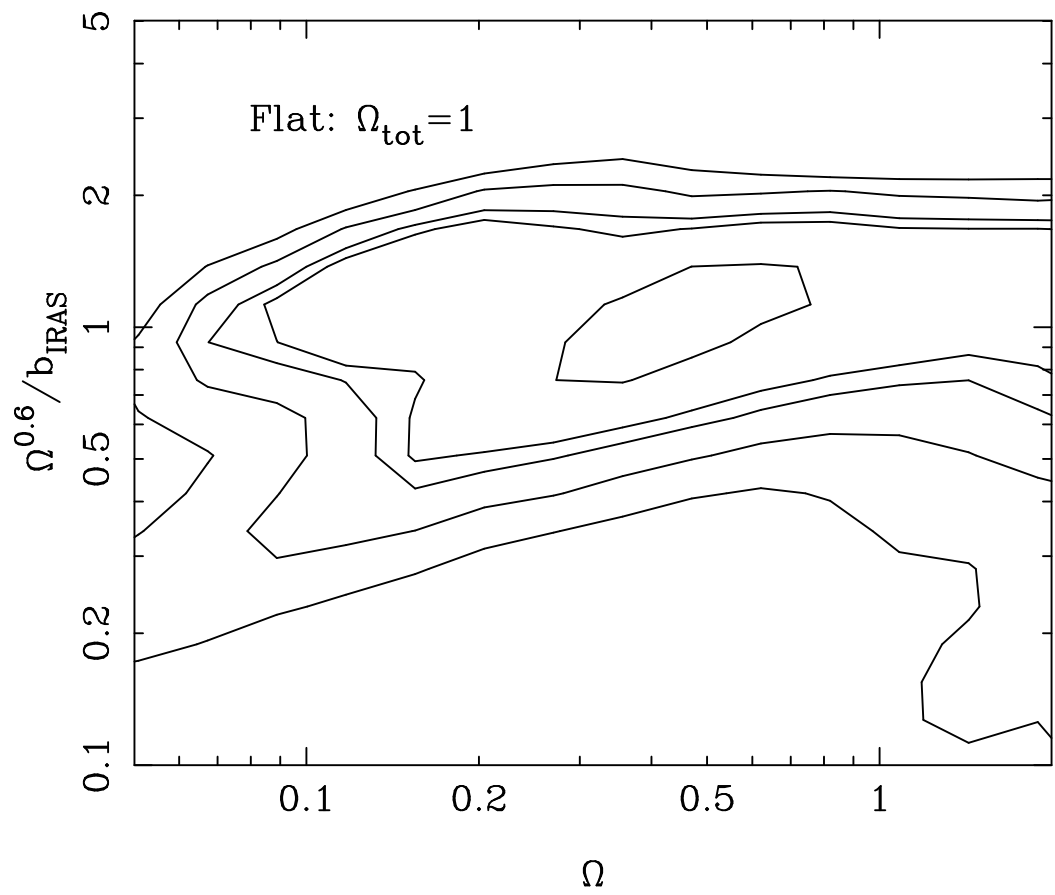
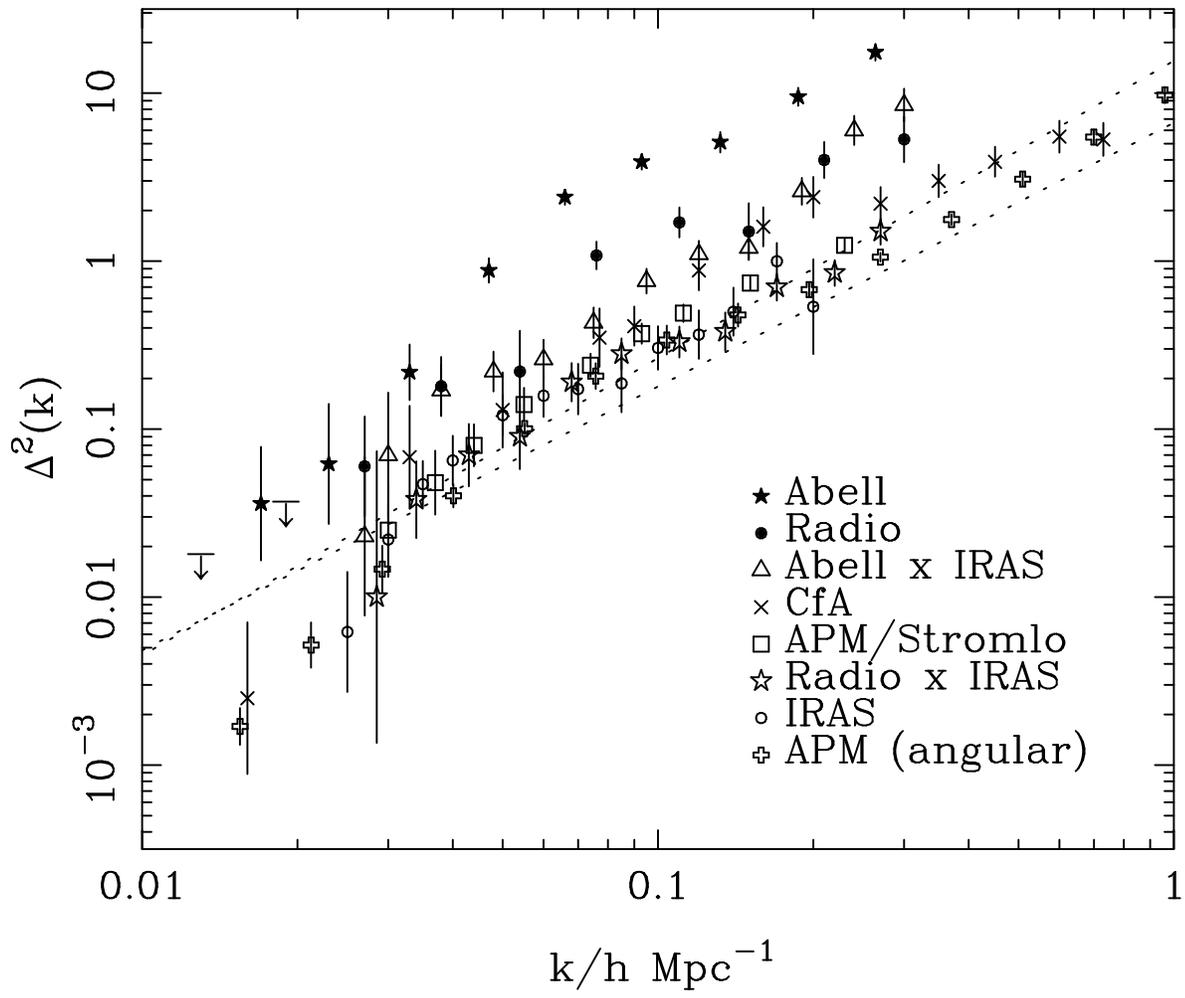
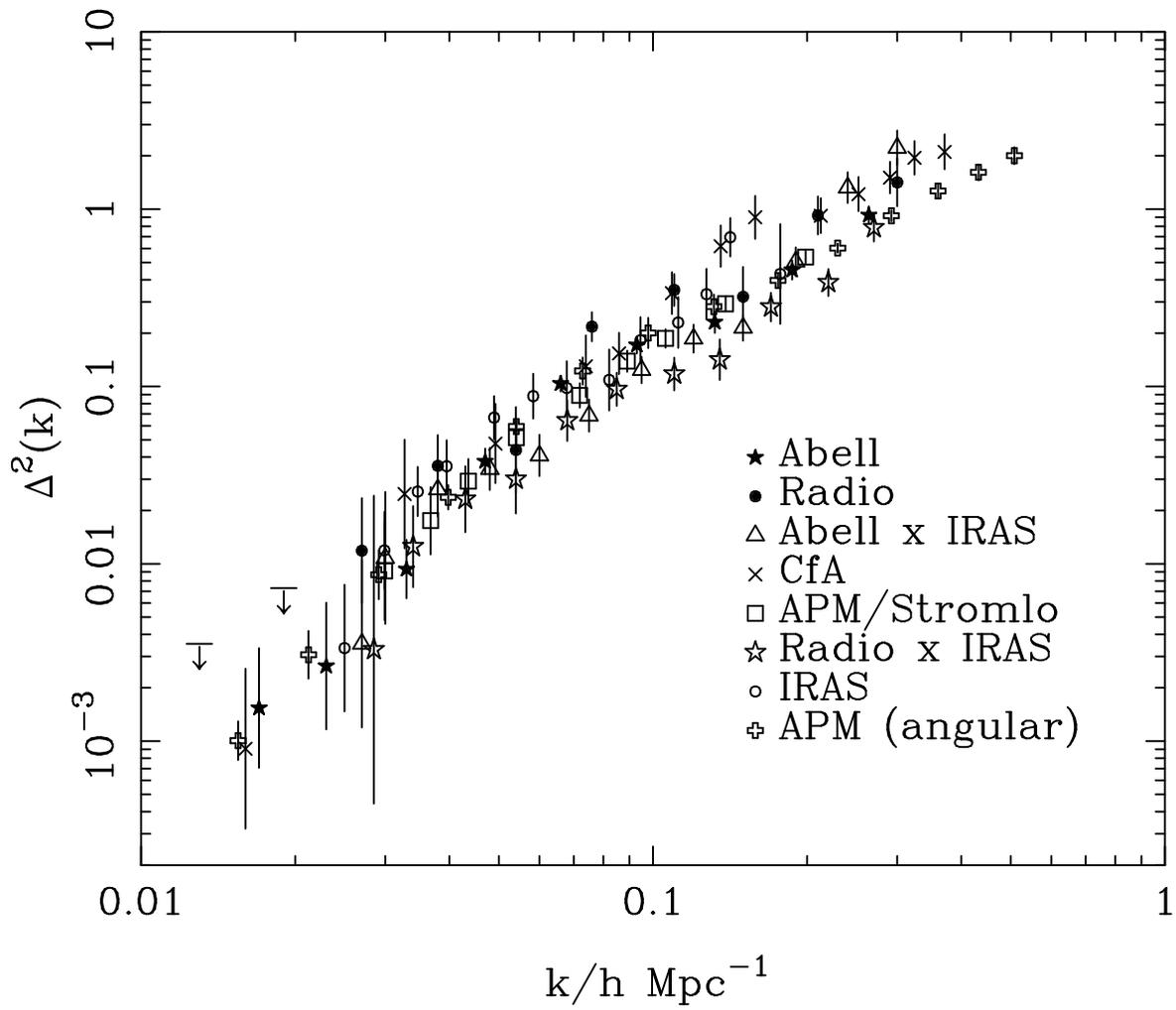
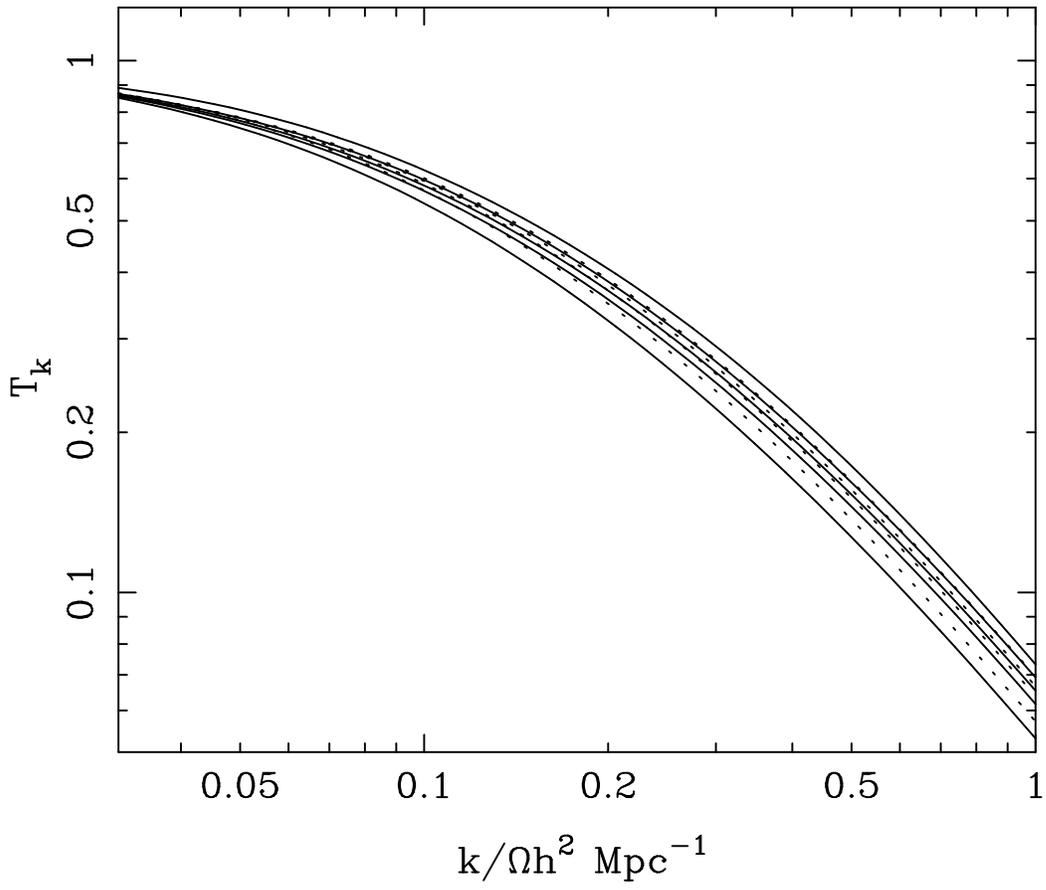
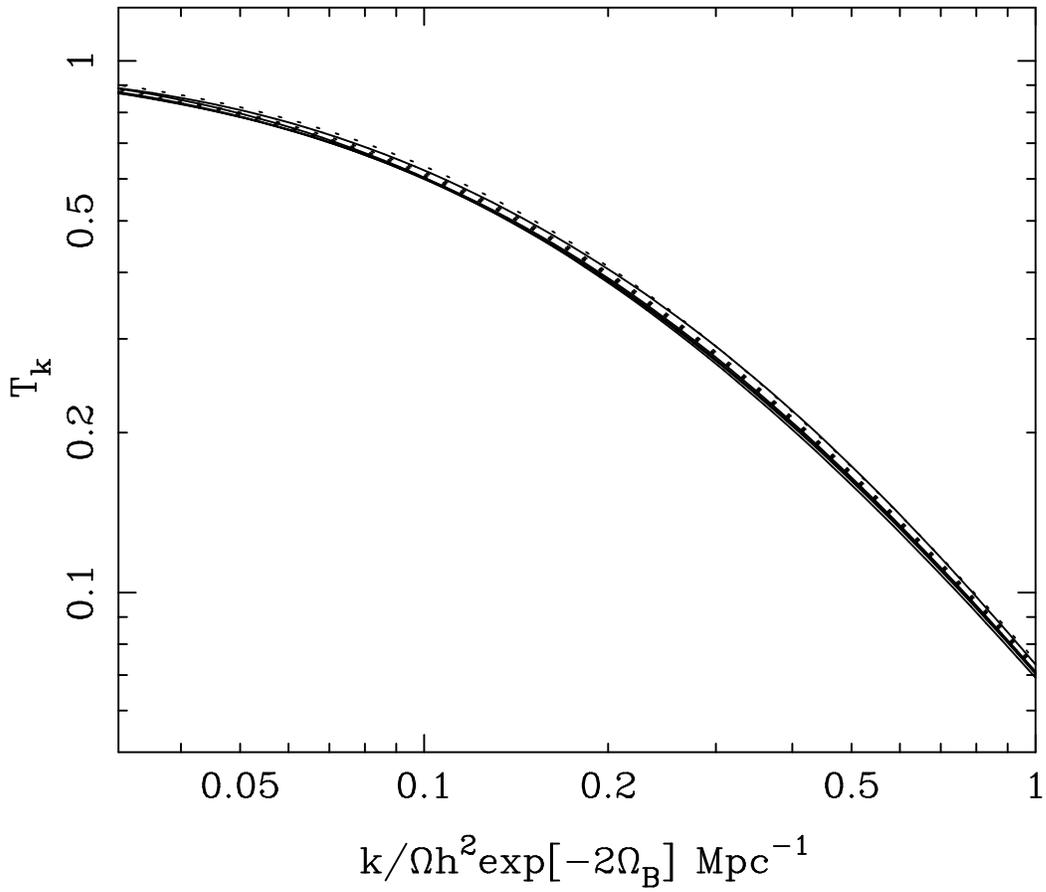
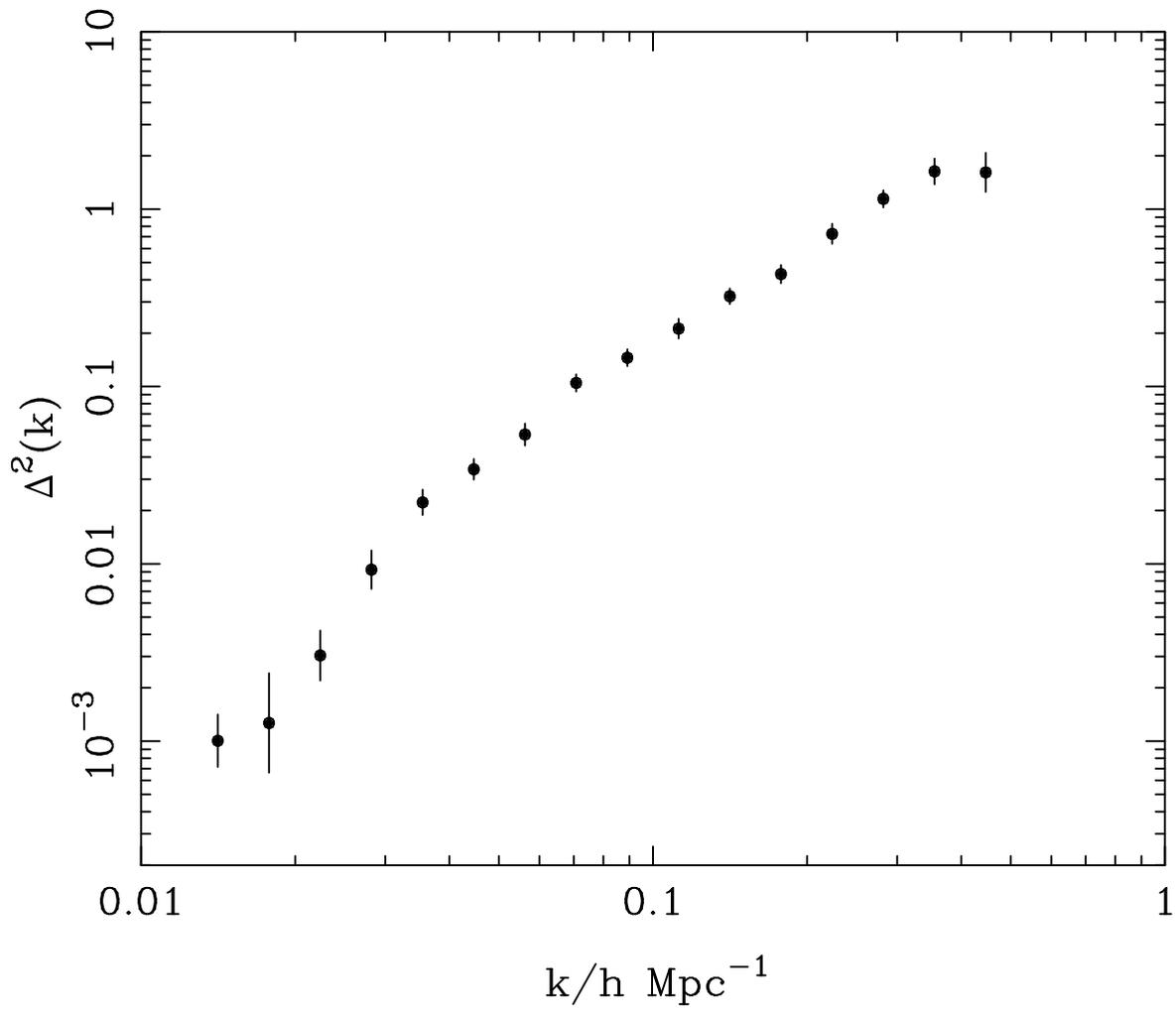
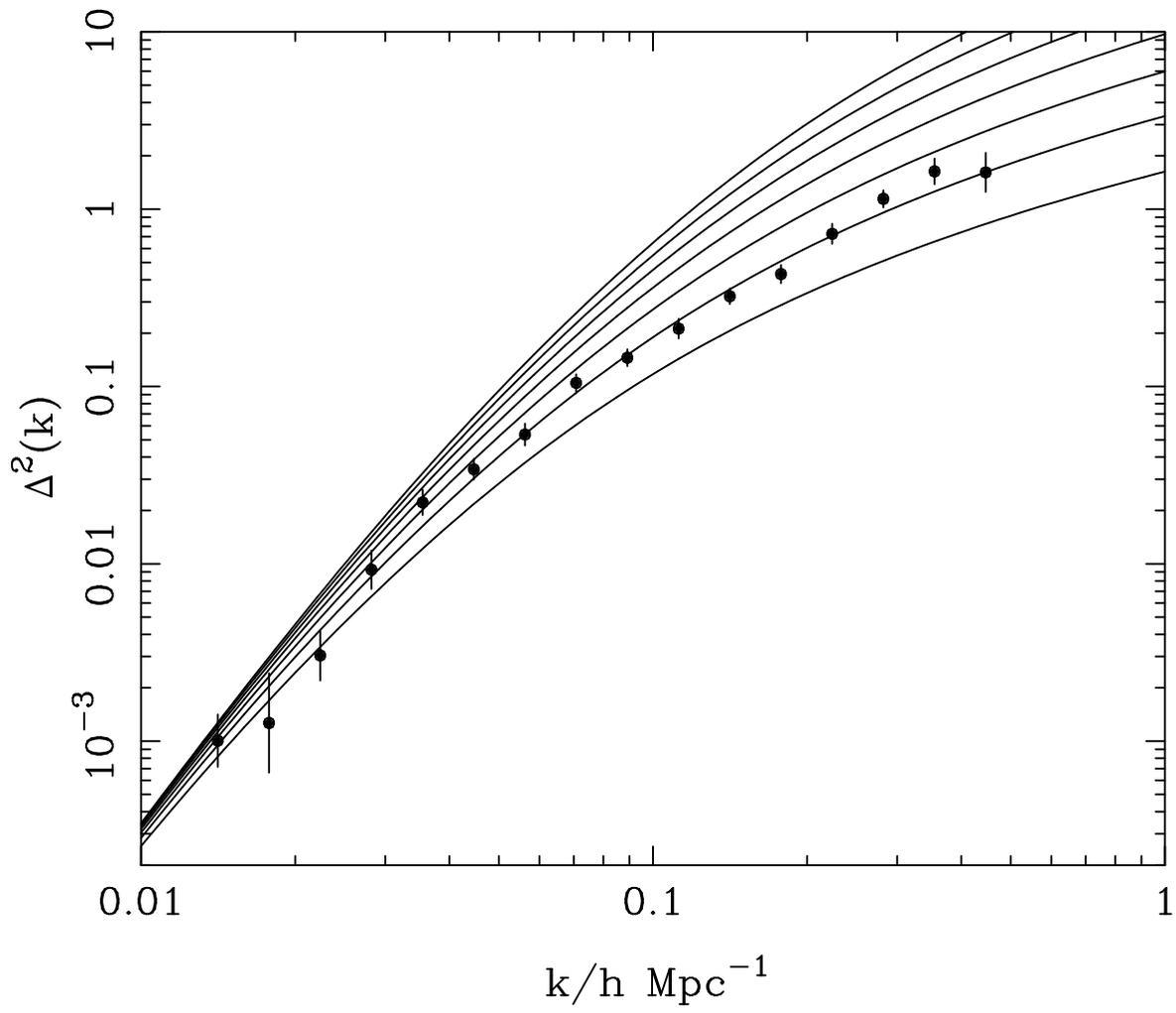
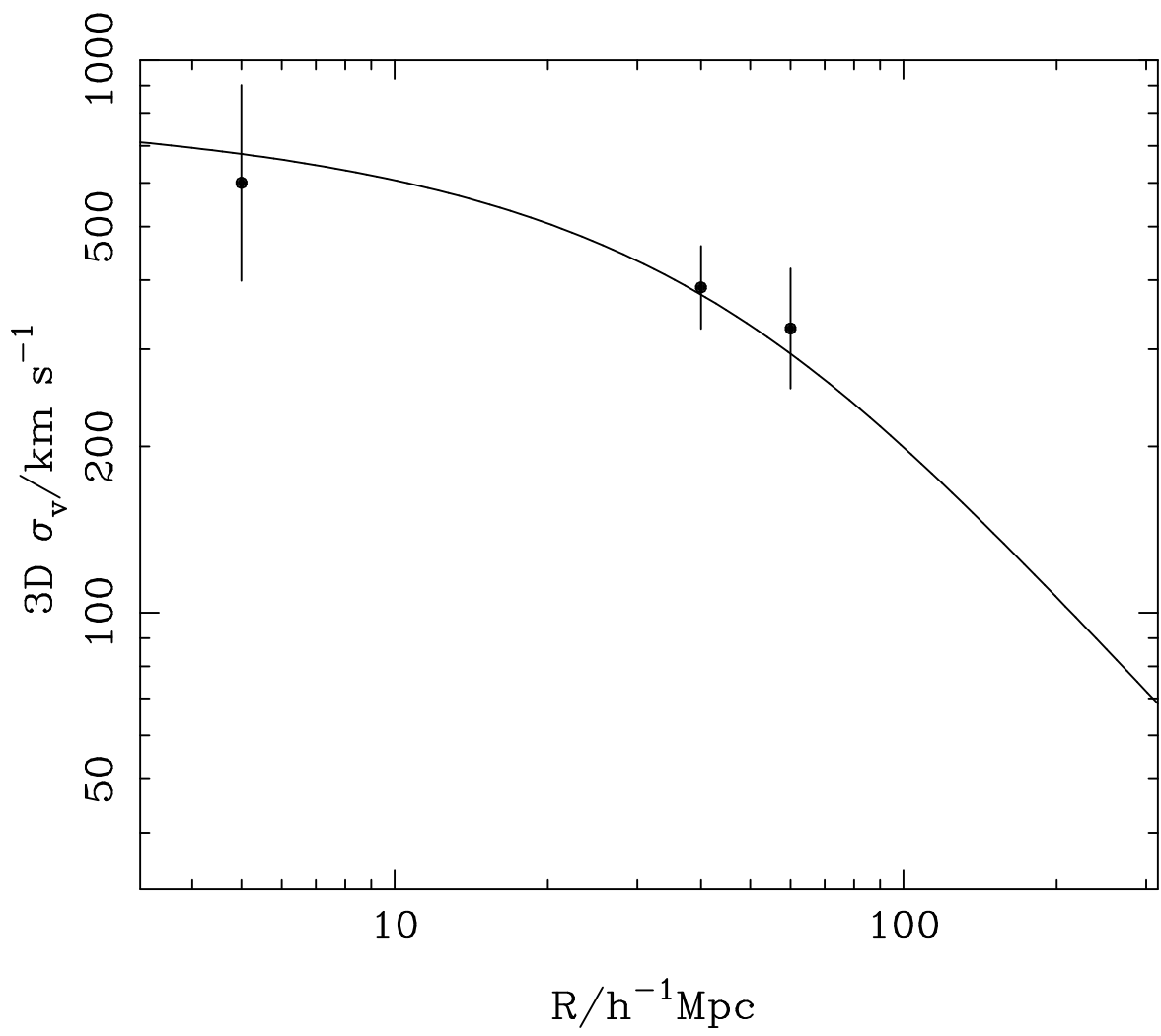
\end{document}